\documentclass[lettersize,journal]{IEEEtran}
\usepackage{cite}
\pdfoutput=1
\usepackage{amsmath,amsfonts}
\usepackage{algorithmic}
\usepackage{array}
\usepackage{textcomp}
\usepackage{stfloats}
\usepackage{url}
\usepackage{verbatim}
\usepackage{graphicx}
\usepackage{subfig}
\usepackage{amsmath}
\def\BibTeX{{\mathrm B\kern-.05em{\sc i\kern-.025em b}\kern-.08em
    T\kern-.1667em\lower.7ex\hbox{E}\kern-.125emX}}
\usepackage{balance}
\usepackage[linesnumbered,ruled,vlined]{algorithm2e}%[ruled,vlined]{
\makeatletter
\let\NAT@parse\undefined
\makeatother
\usepackage{hyperref}  %hyperref still needs to be put at the end!
\begin{document}
\title{Workload Distribution with Rateless Encoding: A Low-Latency Computation Offloading Method within Edge Networks}
\author{Zhongfu Guo, Xinsheng Ji, Wei You, Yu Zhao, Bai Yi, Lingwei Wang
\thanks{Manuscript created August, 2023; Project supported by the National Key Research and Development Program of China (Nos. 2022YFB2902204 and 2020YFB1806607).

Zhongfu Guo, Wei You, Yu Zhao, Yi Bai and Lingwei Wang are with the Department of next-generation mobile communication and cyber space security, Information Engineering University, Zhengzhou 450001, China.

Xinsheng Ji is with the National Digital Switching System Engineering and Technological Research and Development Center, Zhengzhou 450000,China, and he is also with the Purple Mountain Laboratories: Networking, Communications and security, Nanjing 211111, China. (e-mail: ndscjxs@126.com).}}
\markboth{Preprint submitted to Elsevier}%
%\markboth{Computer Network, ~Vol.~ , No.~ ,  ~  }%
{}

\maketitle

\begin{abstract}
This paper introduces REDC, a comprehensive strategy for offloading computational tasks within mobile Edge Networks (EN) to Distributed Computing (DC) after Rateless Encoding (RE). Despite the efficiency, reliability, and scalability advantages of distributed computing in ENs, straggler-induced latencies and failures pose significant challenges. Coded distributed computing has gained attention for its efficient redundancy computing, alleviating the impact of stragglers. Yet, current research predominantly focuses on tolerating a predefined number of stragglers with minimal encoding redundancy. Furthermore, nodes within edge networks are characterized by their inherent heterogeneity in computation, communication, and storage capacities, and unpredictable straggler effects and failures. To our knowledge, existing encoding offloading approaches lack a systematic design and unified consideration of these characteristics. REDC addresses these issues by adaptively encoding tasks, then distributing the workload based on node variations.
In the face of unpredictability failures,  the rateless encoding adaptation provides resilience to dynamic straggler effects. Considering the node heterogeneity and system status, tasks are offloaded to optimal subset "valid" nodes. Load distribution decisions are made based on updates to queuing theory modeling through state feedback. The REDC framework is applicable to EN by improving resource utilization and reducing task sequence execution delays. Experimental results demonstrate our method's effectiveness and resilient performance, maintaining efficacy even in the presence of unstable nodes.
\end{abstract}	
%, optimizing resource utilization within edge networks.我们的自适应编码方案被称为“动态掉队弹性”（DSR），可以处理未指定数量的掉队，从而提供卓越的鲁棒性和弹性。

\section{Introduction}
\IEEEPARstart{T}{he} integration of mobile networks into application scenarios like vehicular networking\cite{javed2022future}, smart factories, and smart homes is gaining attention\cite{navarro2020survey,li20185g}. The 6G mobile edge network (EN)\cite{adhikari20226g}, with its integrated computing and transmission\cite{zhang2022mobile,hassan2019edge}, opens up new possibilities for advanced applications in these areas\cite{yoshizawa2019overview}. EN capitalizes on deploying diverse computing nodes like CPUs, GPUs, FPGAs, and DSPs to notably enhance computational efficiency\cite{liu2019edge,biookaghazadeh2018fpgas}. Such distributed architecture deployment enhances efficiency and scalability. Considering the disproportionate cost-to-benefit ratio of enhancing reliability with a single node\cite{zhen2020distributed}, EN strategically ensure system resilience through multi-nodal structures, computational redundancy, and distributed deployment\cite{guo2023delay,ng2021comprehensive}. The pressing issue of fully leveraging the performance of the EN to cater to workloads demanding low latency and high reliability remains to be solved.

In ENs, nodes\footnote{Throughout this paper, the terms "node" and "worker" are used interchangeably to refer to the computing nodes in the mobile EN.} that fail to respond in a timely manner are typically referred to as "stragglers"\cite{lee2017speeding}. These stragglers could be due to a variety of factors such as resource contention, disk failures, unstable network conditions, and imbalanced workloads\cite{ananthanarayanan2010reining}. Stragglers are typically considered an unavoidable "system noise" in distributed computing systems\cite{reisizadeh2019coded,kim2020coded}, and the unpredictable computational delays they introduce could significantly degrade system performance\cite{dean2013tail}. Coding theoretic techniques have been recently regarded as promising solutions to cope with the challenges in distributed computing\cite{li2017fundamental}. 
For example, coded distributed computing introduce redundant computation, distributed across multiple nodes, so that the entire computational task can be completed from subtasks done by the fastest server, thus alleviating the straggler effects \cite{lee2017speeding, FrigardKRA21}. 
Coded distributed computing has been considered for, e.g., matrix-vector and matrix-matrix multiplication\cite{tauz2022variable,li2016unified,yu2017polynomial,severinson2018block,mallick2020rateless,2020Factored,2018On,dutta2016short,bitar2017staircase,yu2020straggler,kiamari2017heterogeneous,vedadi2021adaptive}, distributed gradient descent \cite{tandon2016gradient}, and distributed optimization \cite{karakus2017straggler}.

Previous investigations have largely concentrated on optimizing the complexities of encoding and decoding \cite{yu2017polynomial, 2020Factored}, with a concurrent trade-off between communication overhead and recovery threshold \cite{2018On, zhang2019model}. A common assumption in these studies is the homogeneous nature of nodes \cite{mallick2020rateless}. 
In practical contexts, the variance among nodes is notable. This is influenced by variances in computational power \cite{lim2020federated}, which can be attributed to factors such as power capacity, workload, and aging, as well as discrepancies in communication delay \cite{asheralieva2021auction}, stemming from the diversity in link bandwidths and stability.
Therefore, integrating node heterogeneity into the design of distributed coded computation is vital for optimizing resource utilization. \cite{kiamari2017heterogeneous} devised a mechanism accommodating such heterogeneity by partitioning, encoding, and distributing a matrix to worker threads. However, their model overlooks the inherent temporal variability of the computing capabilities of the workers.

Newly, \cite{zhang2019model, FrigardKRA21} considered task offloading in ENs, albeit primarily focusing on the balance between coding and communication. Meanwhile, \cite{kim2020coded} centered their research on merging partially completed edge computing tasks for computation recovery. In contrast, our study extends from the unique characteristics of ENs. We contemplate node heterogeneity, random time-varying performance, and encompass the entire lifecycle of offloaded computing execution—including encoding, decoding, communication, waiting for computation, and computation delay. Our approach thereby introduces a dynamic encoding computing offloading mechanism designed for adaptation to heterogeneous and temporally varying node clusters.

\begin{figure*}[!t]
	\centering
	\includegraphics[width=7in]{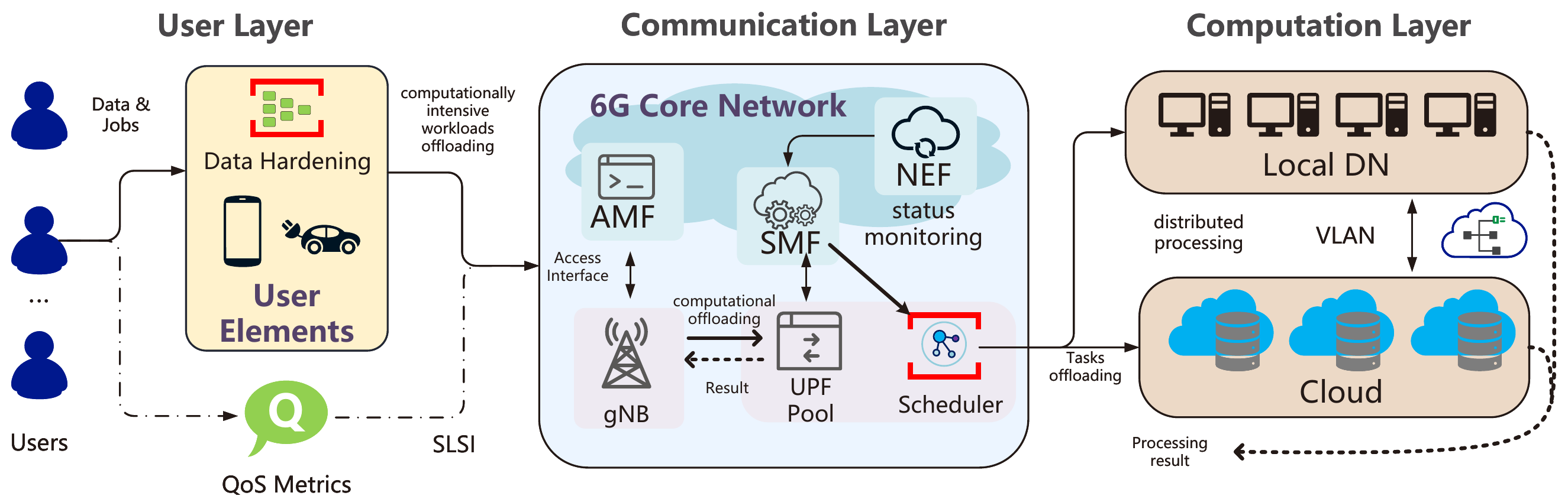}
	\caption{Example scenario for computationally intensive workloads offloading Demand. In our context, User elements continuously generate computationally intensive tasks, which are offloaded to the edge computing nodes (Local DN) and Cloud server for distributed processing. This type of computational offloading is essential to optimize the utilization of resources and improve the overall system performance.}
	\label{ELECS_fig01}
\end{figure*}
Presented herein is a comprehensive computation offloading strategy for the Edge Network, Rateless Encoding Distributed Computing (REDC), engineered to unleash the full computational power of ENs, provide low-latency and high-reliability computing services for workflows. This strategy spotlights matrix multiplication operations, ${\mathbf{C}} = {{\mathbf{A}}^T} \times {\mathbf{B}}$, which are fundamental in various machine learning and data analytics\cite{lee2017speeding}. Taking into account the inherent heterogeneity and time-varying properties of nodes within ENs, REDC implements the M/G/1 queue model for efficient node analysis. This model relies on the response information from the first and second moments of a node's runtime for its updates. 
It empowers the controller to deduce the status of each node and devise efficient scheduling strategies with minimal feedback, which is crucial for optimizing the utilization of EN computational resources. Acknowledging the stochastic characteristics of node failures, REDC integrates an adaptive Rateless Encoder.
It eliminates the need for predefined node failure rates and permits flexible generation of encoding symbols in accordance with the determined scheduling strategy. Furthermore, we introduce a Quick Launch Strategy (QLS). This method reorganizes the sequence of task distribution, thereby curtailing idle periods for nodes and consequently enhancing overall computational efficiency. The contributions of this paper are:
\begin{itemize}
	\item{Proposing REDC, the first EN computation offloading method which integrates data encoding and scheduling in an adaptive manner, optimizing efficiency while ensuring low-latency and high-reliability services, even under conditions of node failures and performance fluctuations.}
	
	\item{In response to the unpredictability of node failure rates, we introduce a feasible strategy that provides resilience for variable stragglers, thereby enhancing the resilience and efficiency of computation offloading.}
	
	\item{We conducted a comprehensive analysis of prior computational characteristics models for nodes, facilitating the construction of distinct node types during simulations.}
\end{itemize}

The structure of this paper is as follows. Section II provides a brief overview of related literature. In Section III, we present our system model. The details of the REDC strategy are discussed in Section IV. Numerical results are shown in Section V. Finally, Section VI concludes the paper and suggests potential directions for future research. Table \ref{tab1} summarizes the important variables used in this paper.

\begin{table}
	\begin{center}
		\caption{Table Of Basic Notations.}
		\label{tab1}
		\begin{tabular}{| l | l |}
			\hline
			Symbol & Definition \\
			%of filter & $e_m$ \\
			\hline
			${GF}_{(q)}$ & The Galois Field of size $q$\\
			\hline
			${\mathbb{E}}$  & The expectation value\\
			\hline
			$N_I$ &  The total number of available computing nodes\\
			\hline
			$\mathcal{I}$ & A valid set of computing nodes\\
			\hline
			$m,k$ & The count of divisions applied to matrices \textbf{A} and \textbf{B} \\
			\hline
			$\mathcal{N}$ & The total number of computational units (CUs) output by the encoder\\
			\hline
			$N$ & The count of CUs required to finalize decoding\\
			\hline
			$\Gamma$ & The ratio of redundancy per job\\
			\hline
			%${\varpi _i}$ & The number of tasks assigned to the $i$-th worker per job iteration\\
			%\hline
			$w_i$ &  Identifier for the $i$-th worker\\
			\hline
			${{\varpi_{i}}}$ & The task service time reported by the $i$-th worker\\
			\hline
			$\texttt W_{i}$ & Tasks assigned to the $i$-th worker\\
			\hline
			$\upsilon$ & The  arrival rate of Jobs\\
			\hline
		\end{tabular}
	\end{center}
\end{table}

\section{Related Work and Motivation}
In the epoch of ubiquitous connectivity and big data, the explosion of terminal data has outstripped the projections of Moore's Law. Sixth-generation (6G) Edge Networks (ENs) amalgamate wide-coverage mobile networks with distributed computing nodes\cite{adhikari20226g}. This integration is purposed to facilitate ultra-reliable, low-latency communication and computation services at the network's edge\cite{tomkos2020toward}, in effect, mitigating latency and energy consumption\cite{zhang2022mobile}. 
A representative 6G Edge Network scenario is illustrated in Figure.\ref{ELECS_fig01}. It supports cloud-edge-device collaborative computing. User Elements (UE) with limited computing power have computing-intensive business flow computing requirements. The mobile Edge Network reads the user's Service Level Subscription Information (SLSI) based on Edge computing nodes, or cloud servers provide computing offloading services. 
The Session Management Function (SMF) of the Core network can establish a Protocol Data Unit (PDU) session to deliver workload\cite{3gpp.23.502}. Meanwhile, the Network Exposure Function (NEF) can access node status and adjusts the allocation of computing power\cite{3gpp.36.331}. Consequently, the EN's offloading decision can be executed at the core network's control plane. We pay attention to the efficient operation of workload, establish computing redundancy through coding, and distribute computing tasks to appropriate computing nodes through appropriate scheduling strategies. The work of this paper is marked as a red box in Figure.\ref{ELECS_fig01}.

The functional components of EN code computing offload considered by REDC have five parts, as shown in Figure.\ref{ELECS_fig02}a: the controller acts as a strategy center, receiving status feedback from computing nodes and decoders, and issuing offload coding policies, scheduling policies, and purging instructions; Encoder adds redundancy to the computing tasks through a suitable coding method; Scheduler: based on the status feedback of the computing nodes, offloads the tasks to the nodes of the EN; worker: caches the assigned tasks and perform calculations; Decoder: decodes encoded calculation tasks.
\subsection{Coded Computing} 
Within a distributed computing system, the presence of stragglers — inefficient nodes that slow down overall computation — is a persistent issue that cannot be entirely eradicated from the computing cluster. A plausible approach to counter stragglers involves amplifying computational redundancy through replication \cite{ananthanarayanan2013effective,shah2015redundant,wang2014efficient,gardner2015reducing,chaubey2015replicated,lee2016scheduling,joshi2017efficient}. However, the simplistic replication approach comes with a trade-off, the escalating communication and computational overheads.

\begin{figure}[!]
	\centering
	\includegraphics[width=3.5in]{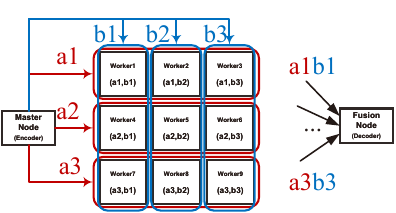}
	\caption{Illustration of coded distributed computing: 2D-MDS Code ($(3,2)^2$ Product Code)\cite{lee2017high} in an example with $N = 9$ workers that can each store half of $\mathbf{A}$ and half of $\mathbf{B}$, where $\text a = [\text a1,\text a2,\text a3] ^{\text T}= [\mathbf{A}_1,\mathbf{A}_2,\mathbf{A}_1+\mathbf{A}_2] ^{\text T}$,$\text b = [\text b1,\text b2,\text b3]= [\mathbf{B}_1,\mathbf{B}_2,\mathbf{B}_1+\mathbf{B}_2]$.}
	\label{ELECS_fig05}
\end{figure}

In a bid to simultaneously tackle the straggler problem and manage computational and communicational load, coded computing\cite{tauz2022variable,li2016unified,yu2017polynomial,severinson2018block,mallick2020rateless,2020Factored,2018On,dutta2016short,bitar2017staircase,yu2020straggler,kiamari2017heterogeneous,vedadi2021adaptive} have emerged as a compelling answer to the complexities of distributed computing. The defining feature of these techniques is their capability to recover the final result from a subset of partial results. Figure.\ref{ELECS_fig05} illustrates a Product Code \cite{lee2017high} method, which partitions the computational matrix into $m=2$ divisions, encoding it into $\sqrt N =3$ parts. The recovery threshold here is: ${K_{{\rm{product}}}} = 2\left( {m - 1} \right)\sqrt N - {\left( {m - 1} \right)^2} + 1$. Another method, the Polynomial Code\cite{yu2017polynomial}, severs the link between the recovery threshold and the number of nodes, making the threshold exclusively dependent on the splitting numbers $m$ and $n$ of matrices $\textbf A$ and $\textbf B$. In their work, Dutta \textit{et al.}\cite{2018On} delved into the interrelationship between node computation, communication overhead, recovery threshold, and matrix splitting numbers, proposing a lower recovery threshold of ${K_{{\rm{MatDot}}}} = 2m- 1$. Such studies typically employ an erasure model where the maximum number of anti-stragglers is pre-set, and the recovery threshold is defined as the minimum count of worker threads needed to yield results for successful decoding, a concept termed as the certainty threshold \cite{2020Factored}. While the fixed recovery threshold scheme may be found wanting in the face of time-varying nodes, the ACM$^2$\cite{vedadi2021adaptive} proposes an automated selection method that consolidates multiple coding strategies, albeit at the cost of an increase in system complexity.

\subsection{Rateless Code}
The idea of rateless encoding was originally proposed by Luby \textit{et al.}\cite{luby2002lt}, and it is an encoding method with an unfixed code rate. The generation matrix is randomly generated according to the degree distribution, and the rate of rateless encoding is not determined. The decoding threshold, whose recovery probability is related to the number of encoded symbols received by the decoder, the larger the value, the higher the probability, which is called the probabilistic threshold. Taking the first practical fountain code LT code as an example, the coding parameters of the fountain code can be expressed as LT(K, $\Omega(x)$), K represents the number of source symbols participating in the coding, and $\Omega(x)$ represents the degree distribution of the output symbol. Specifically, ${{\Omega _d}}$ stands for the likelihood that the degree value is d. Typically, N symbolizes the number of encoded output symbols, where N=(1+$\epsilon$)K. Here, $\epsilon$ stands for the redundant coefficient, also known as the decoding overhead. By selecting an appropriate degree distribution, the overhead $\epsilon$ tends towards zero as $K$ approaches infinity.

Fountain codes have good adaptability in distributed transmission systems, and have also been studied in distributed computing. Anton \textit{et al.}\cite{FrigardKRA21} designed a coding scheme for the EN, consisting of the concatenation of a rateless code and an irregular-repetition code, taking into account decoding delays. Mallick \textit{et al.}\cite{mallick2020rateless} considered the adaptive coding mechanism of heterogeneous time-varying resources. Rateless coding has the ability to cope with unpredictable node failures in the system, and the decoding overhead is negligible when the code length is long. Considering that the reliability and resilience of mobile edge computing will become the first indicator restricting its development in the field of industrial Internet and other fields, it is completely acceptable to pay for the cost of improving the resilience and reliability of the system in the current rich media era. At the same time, this kind of reliability improvement through multi-node computing redundancy is more cost-effective than single-node.

\subsection{motivation}
In the paradigm of Coded Distributed Computing (CDC)\cite{ng2021comprehensive}, computation workloads are encoded and subsequently offloaded to the EN, a concept we refer to as Coded Computing Offloading (CCO). The goal of CCO is to achieve enhanced latency and precision benefits. Prior works have contemplated decoding delays and trade-offs between computation and communication \cite{FrigardKRA21, kim2020coded, zhang2019model}. The mobile EN shoulders the task of handling computing-intensive terminal workloads, offering computation offloading services via nearby distributed computing resources. Primarily, the subsequent key factors are taken into consideration:

A) Load Balancing: Given the discrepancies in computation power, storage, and energy resources among edge devices, efficient task and data distribution across the network is imperative for ensuring load balance and optimizing overall performance.

B) Resilience: Nodes in a mobile edge network might intermittently enter or exit the network, or undergo variations in resource status. Thus, the design of encoding computation solutions must withstand such dynamism. Concurrently, given the unpredictability of node failure rates, the task offloading scheme must incorporate flexibility to ensure reliable task execution.

C) Total Latency Considerations: The total execution time for computation offloading incorporates task encoding, data uplink transmission, task distributed computation, result return, and decoding. The task can only be submitted after all these processes are completed, hence the need for comprehensive consideration of the actual execution procedure.

D) Task Characteristics: We consider the random arrival of tasks. Moreover, we commence specific task analysis with matrix multiplication operations. This approach offers greater universality compared to matrix-vector operations or solely offloading vectors.

We consider providing a reliable and stragglers-resilient CCO service. This service takes into account the heterogeneity of edge network computing nodes, time-varying computing performance, and unpredictable node failure rates. Based on these considerations, we propose a set of coding computing strategies. The next section will conduct a modeling analysis for system elements.

\begin{figure*}[!t]
	\centering
	\includegraphics[width=7in]{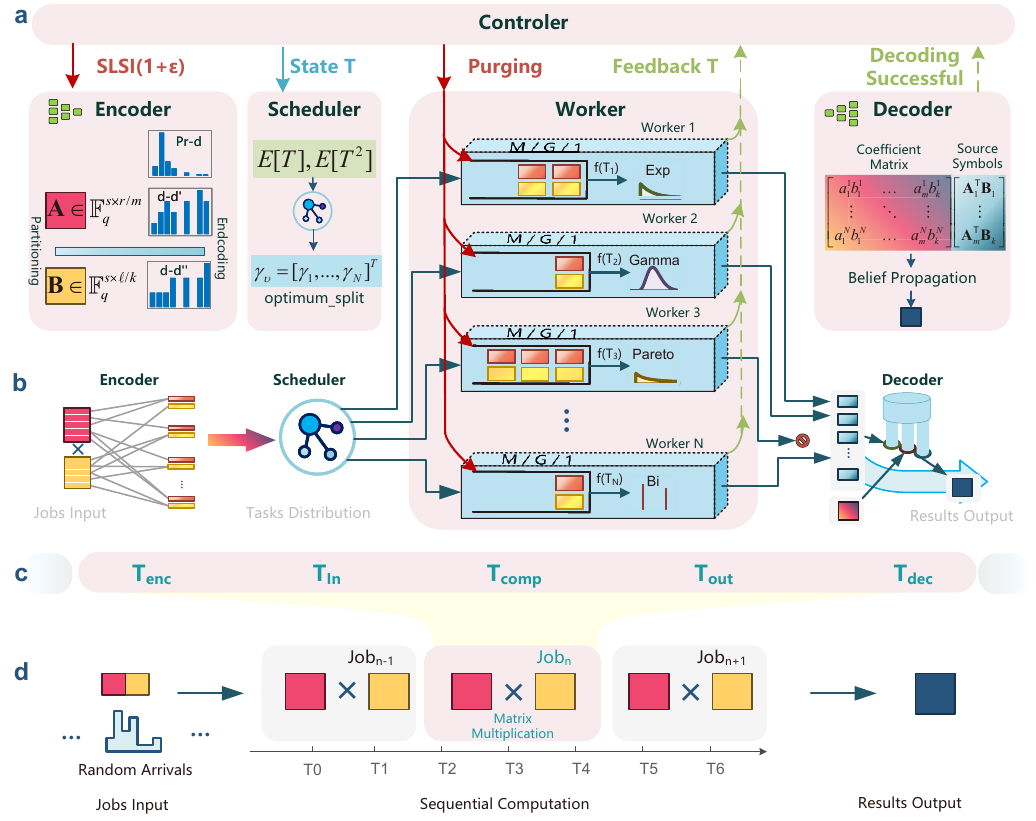}
	\caption{Overview of the Rateless Encoding Distributed Computing (matrix multiplication) framework. a. The system includes controllers, encoders, schedulers, workers, and decoders; b. Focus on matrix operations and show the operation process; c. Consider calculation execution time including encoding time, data upload time to computing nodes, and operation time. Calculation result return time, decoding time;	d. The computing tasks we consider are executed sequentially, and the computing tasks of the workload arrive randomly.}
	\label{ELECS_fig02}
\end{figure*}

\section{System Model}
The edge network harbors a wealth of heterogeneous computing resources and carries out workloads through a distributed computing framework. We focus on key functional entities, depicted in Figure\ref{ELECS_fig02}.a, encompassing the master node (task requestor), encoder, scheduler, computational node, controller, and decoder. These functional modules are abstracted from the distributed coding computation within the edge network, and they do not correspond entirely one-to-one with physical components. This section offers a detailed exploration of the system model, primarily discussing it from both the functional (processing flow) and temporal overhead (time characteristics) perspectives.
\subsection{Workload Model}
Figure.\ref{ELECS_fig02}.d illustrates the timeline of executing iterative operations sequentially from the master node's perspective. The master node systematically receives inputs from low-latency constrained workloads. We model the arrival rate of these Jobs using a Poisson process with parameter $\upsilon$. Here, $N(t)$ represents the count of Jobs arriving within $t$ time slot, where $t \in { 1,...,T} $. The process adheres to the following probability distribution:
\begin{equation}
	{\mathbb{P}} (N(t))= e^{-\upsilon t} \frac{(\upsilon t)^{n}} {n!}
\end{equation}
where the mean number of Jobs arriving per slot is $\upsilon$.

In the distributed computing offloading paradigm,the master node splits the workload, i.e, matrix multiplication, into smaller sub-matrices and distribute them to the Edge Network (EN) for parallel execution. UE handles computation workloads, termed "Jobs," that arrive at random. Each Job signifies a matrix multiplication request, which is denoted as ${{\mathbf{A}}^T} \times {\mathbf{B}}$, where $A\in \mathbb{F}^{s \times r}$, and $B\in \mathbb{F}^{s \times \ell}$. The most atomic unit of these operations is referred to as a Compute Unit (CU), representing a singular row(vector) multiplication operation. A Task comprises a certain number of CUs.

The computation results corresponding to matrices A and B can be represented in the form of matrix $\mathbf {C}$, whose elements are calculated as: 
$\mathbf {C}_{ij} = \mathbf {A}_{x}^{\mathsf {T}}\mathbf {B}_{y}$. Here, $x \in [m]$ and $y \in[k]$. Thus, calculation of matrix $\mathbf {C}$ translates to computing $m*k$ sub-matrices.
For the worker $w_i$, $i \in {N_I}$  involved in matrix multiplication operations, whose computation tasks $\texttt {W}^{i}_{\mathbf {A }} \times \texttt {W}^{i}_{\mathbf {B}}$ represent subsets of [m] and [k] respectively.  
We summarize earlier research and define them as:
\begin{equation} 
	{\mathbf {\tilde {A}}_{i}}^{\mathsf {T}}=\sum _{x \in [m]}a_{x}^{i}\mathbf {A}_{x}^{\mathsf {T}}, \quad \mathbf {\tilde {B}}_{i}=\sum _{y \in [k]}b_{y}^{i}\mathbf {B}_{y}
\end{equation}
Each encoding scheme corresponds to a coefficient vector, $\vec {a}^{i}=[ a_{1}^{i}, a_{2}^{i}, \cdots, a_{m}^{i}]$ and $\vec {b}^{i}=[b_{1}^ {i}, b_{2}^{i}, \cdots, b_{m}^{i}]$, essentially the encoded vector. Each worker calculates $\mathbf {\bar {C}}^{i}={\mathbf {\tilde {A}}_{i}}^{\mathsf {T}}\mathbf {\tilde {B }}_{i}$ and returns $\mathbf {\bar {C}}^{i}$ to the master node.

Decoder collects $\mathbf {\bar {C}}^{i}$ returned from workers, deemed non-stragglers. The operation $\mathbf {\bar { C}}^{\mathsf {T}}=[\mathbf {\bar {C}}^{1},\mathbf {\bar {C}}^{2},\ldots, \mathbf {\bar {C}}^{N_i}]$ constitutes the encoding operation, whereas the process of deducing the matrix $\mathbf {C}$ from the gathered computation results is the decoding operation. We collectively refer to a set of codec scheme as an encoding computation. Meanwhile, the computation task distribution mechanism and the encoding scheme are collectively referred to as the coding-computation offloading scheme.
In summary, this subsection presents the proposed workload model for distributed computing offloading. This model lays the groundwork for further analysis and strategy development.

\begin{algorithm}[!t]
	\caption{Distributed Computing Rateless Encoder}
	\label{alg:algorithm1}
	\KwIn{$\textbf{A}$, $\textbf{B}$, $m$, $k$, $\mathbf{\Omega}(x)$}
	\KwOut{${\mathbf{\tilde A}}$, ${\mathbf{\tilde B}}$}
	$\textbf{A}_\textbf{M} = [\textbf{A}_{r/m \times s}^\textbf{1},\textbf{A}_{r/m \times s}^\textbf{2},...,\textbf{A}_{r/m \times s}^\textbf{m}] \leftarrow \textbf{split}[\textbf{A}_{r \times s}]$\;
	$\textbf{B}_\textbf{K} =[\textbf{B}_{s \times \ell/k}^\textbf{1},\textbf{B}_{s \times \ell/k}^\textbf{2},...,\textbf{B}_{s \times \ell/k}^\textbf{k}] \leftarrow \textbf{split}[\textbf{B}_{s \times \ell}]$\;
	\While{$\{ \sim decoding \_ success_{tag}\}$}
	{
		$\textbf{Init}[\textbf{G}',\textbf{G}'',\textbf{Z},{\mathbf{\tilde A}},{\mathbf{\tilde B}}]$\;
		\For{${i} = 1$ to $\aleph$}
		{
			$d \leftarrow$ $\textbf{RandomSample}$[$\mathbf{\Omega}(x)$]\;
			$\{\mathcal {D}\} \leftarrow$  Compute all divisors of $d$\;
			$d' = \mathop {\mathrm {argmin}} _{x \in \{\mathcal {D}\}} |x-\frac {d}{x}|$,
			$d'' = d/d'$\;
			$\{\vec a,\vec b\} \leftarrow$ $\textbf{RandomVector}$\{[$d',m$], [$d'',k$]\}\;
			$\textbf{G}' \leftarrow \textbf{append}[\textbf{G}',\vec a]$, $\textbf{G}'' \leftarrow \textbf{append}[\textbf{G}'',\vec b]$\;
			$\vec z = \vec a \odot \vec b$\;
			$\textbf{Z} \leftarrow \textbf{append}[\textbf{Z},\vec z]$\;
		}
		${\mathbf{\tilde A}} \leftarrow \textbf{G}_{\mathcal{N} \times m}' \textbf{A}_\textbf{M}$\;
		${\mathbf{\tilde B}} \leftarrow \textbf{G}_{\mathcal{N} \times k}'' \textbf{B}_\textbf{K}$\;
		\textbf{return} ${\mathbf{\tilde A}}$, ${\mathbf{\tilde B}}$, $\textbf{Z}$\;
		$\textbf{Wait}$ $\mathcal{T}$\;
	}
\end{algorithm}

\subsection{Encoding Model}
As illustrated in Figure.\ref{ELECS_fig02}.b, the process begins with the encoding of the Job. In an effort to integrate fountain codes within matrix multiplication operations, we propose an encoder.\ref{alg:algorithm1} tailored to distributed computing. We regard the operation $\mathbf {A}_{i}^{\mathsf {T}}\mathbf {B}_{j}$
as a Computational Unit (CU), where $i \in [m]$ and $j \in [ n]$. Employing an appropriate encoding technique, the matrix multiplication problem is then transformed into an operation on each encoded symbol ${\mathbf{\bar A}}$ and ${\mathbf{\bar B}}$.

It becomes evident that in traditional single-channel distributed scenarios, the degree value $d$, randomly generated according to $\Omega(x)$, is not directly applicable for encoding matrix operations. As per Algorithm \ref{alg:algorithm1}, we offer a concise description of the encoding process. Initially, the encoder splits the input matrix, as depicted in lines 1 and 2. This parallels the process of defining the code length $K$ in conventional LT codes. 
The columns of matrices $\textbf{A} \in {\mathbb{F}}_{s \times r}$ and $\textbf{B} \in {\mathbb{F}}_{s \times \ell}$ are subdivided into $m$ and $k$ portions respectively, denoted as: $ \mathbf {A}=\left [{\mathbf {A}_{1},\mathbf {A}_{2},\ldots, \mathbf {A}_{m}}\right], {\;\text { and }}\; \mathbf {B}= \left [{ \mathbf {B}_{1}, \mathbf {B}_{2},\ldots, \mathbf {B}_{k}}\right]$. Here, we set $K=m*k$, which is the count of essential CUs requiring computation. The selection criteria for $m$ and $n$ will be elaborated upon in the subsequent discussion. For the present, we treat them as input parameters for the encoder.

The subsequent encoding phase involves constructing the generating matrix $\textbf{G}$. Based on $\Omega(x)$, the encoder arbitrarily selects the base degree value $d$ where $d \le K$. The probability of selection is $\mathbb {P} \{Deg = d\} = \Omega_{d}$. The coding degree values $d'$ and $d''$ for matrices $\textbf{A}$ and $\textbf{B}$ need to be defined, taking into account the solvability of coded symbols. As detailed in line7, we select $d'$ from $\{\mathcal {D}\}$,  where $\{\mathcal {D}\}$ represents the set of divisors of $d$, select $Deg=d'$ ensures the smallest  $|Deg-\frac {d}{Deg}|$. Given that the total degree of matrix multiplication equals $d$, we have $d''=d/d'$.

The process that follows entails the generation of encoded vectors $\vec a\in \mathbb{F}^{1 \times m}$ and $\vec b\in \mathbb{F}^{1 \times n}$, with Hamming weights $d'$ and $d''$, respectively, these vectors consist of elements '0' and '1'. One possible approach is to assign the positions of the $d'$ '1's in the vector by generating a Gaussian distribution over all positions and selecting the top $d'$ for '1's. 
In practical applications, the creation of generating vectors and encoding symbols  occur simultaneously. For the sake of logical explanation of the encoding operation, we refer to lines 10 and 11.

To facilitate decoding, we also need to establish the coefficient matrix $\textbf{Z}_{ N \times m*n}$, defining $\vec z$ as the row of  $\textbf{Z}$, with $z \in {0, 1}^{mk}$. We set $\vec m = \vec a \odot \vec b$, where $\odot$ symbolizes the Cartesian product. As depicted in the Decoder of Figure.\ref{ELECS_fig02}.a, the coefficient matrix is directly applied to the decoding operation, which will be discussed in next subsection.

It is of importance to note that in real-world applications, the generation of encoded symbols by fountain codes is not unbounded. If after the initial transmission of $\mathcal{N}$ encoded symbols, successful decoding is not achieved even after waiting $\mathcal{T}$ time slots, we proceed with the supplementary transmission of the subsequent $\mathcal{N}$ encoded symbols. This process continues until either successful decoding is accomplished or the maximum rounds is reached. This mechanism is captured in the while loop from lines 3 to 16.

\begin{figure}[!t]
	\centering
	\includegraphics[width=3.5in]{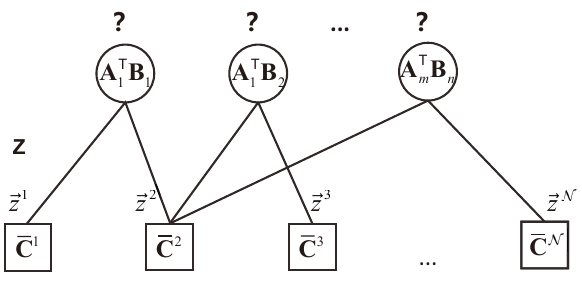}
	\caption{bipartite graph.}
	\label{ELECS_fig03}
\end{figure}

\subsection{Decoding Model}
In consideration of stragglers in the edge network, it is assumed that the decoder receives $N$ computational results. Given the previously defined encoding scheme, the crucial elements in the decoding analysis can be represented as follows:
\begin{align} 
	\begin{bmatrix} \mathbf {\bar {C}}^{1} \\ \mathbf {\bar {C}}^{2} \\ \vdots \\ \mathbf {\bar {C}}^{N} \end{bmatrix} 
	= 
	\underbrace{ \begin{bmatrix} 		
			{\vec z_{}^1}&\;{\vec z_{}^2}&\; \cdots &\;{\vec z_{}^N}
			 \end{bmatrix}^{\mathbf{T}} }_{\mathbf {\mathbf {Z}}} 
	\begin{bmatrix} \mathbf {A}_{1}^{\mathsf {T}}\mathbf {B}_{1} \\ \mathbf {A}_{1}^{\mathsf {T}}\mathbf {B}_{2} \\ \vdots \\ \mathbf {A}_{m}^{\mathsf {T}}\mathbf {B}_{n} \end{bmatrix}
\end{align}
The transition from the encoding to the decoding process necessitates a shift in focus to the coefficient matrix.
Successful decoding requires the collection of an adequate number of encoding packets by the decoder, which ensures the coefficient matrix retains column full rank.

To further illustrate the decoding operation, we employ a Belief Propagation(BP)\cite{casado2007informed} decoding algorithm visualized through a bipartite graph, as shown in Figure.\ref{ELECS_fig03}: Circle nodes, representing the source/input symbol set ${\mathbf{A}}_{i}^T \times {\mathbf{B}}_{j}^{}$ where $i \in [m]$ and $j \in [n]$; the square node represents the output symbol set ${\mathbf{\bar C}}^{n}$ for $n \in [\mathcal N]$. That is, the decoding calculation transforms solving ${\mathbf{\bar C}}^{n}$ into solving ${\mathbf{A}}_{i}^T \times {\mathbf{B}}_{j}^{}$.

The iterative operation of the BP algorithm is central to our proposed decoding model. The process begins by identifying an output node with a degree of one, in Figure.\ref{ELECS_fig03}: ${\mathbf{A}}_{1}^T {\mathbf{B}}_{1}^{} = {\mathbf{\bar C}}^{1}$. Once such a node is located, the source node linked to it is updated, and all edges associated with this refreshed source node are subsequently eliminated. This iterative method persists until all source symbols have been recuperated, signifying successful decoding. Alternatively, the absence of a symbol node with a degree of one results in the termination of the decoding algorithm, indicating decoding failure.

This proposed decoding model, paired with our encoding model, forms the foundation of our approach towards integrating fountain codes within matrix multiplication operations in distributed computing environments.

\subsection{Worker Model}
In distributed computing within edge networks, the process is divided into two primary stages: data transmission\footnote{This includes the transmission time for offloading tasks to computing nodes and returning calculation results, that is, the time spent in preparation, including transmission time for offloading tasks to computing nodes and returning calculation results, excluding the system's own calculations within a single calculation process.} and task execution. A task can fail for two main reasons: 1) if a worker or link malfunctions, leading to a 'Fault' state, or 2) if there are transmission delays or performance degradation, leading to a 'Late' state. Stragglers, workers that exceed the task's time limit, behave like system noise, impacting the performance of distributed computing. By studying timing characteristics during task execution, we can model or analyze the straggler behavior in distributed systems. Computational load and performance are always included in the eigenvalues of the distribution used to describe computation time. As illustrated in Figure.\ref{ELECS_fig02}.d's worker, various distribution models such as the Pareto distribution \cite{joshi2017efficient}, the Markov binomial distribution \cite{yang2019timely}, and the Weibull distribution\cite{reisizadeh2019coded} have been employed to represent computation time's distribution. However, the delay exponential distribution is the most commonly adopted \cite{mallick2020rateless,dutta2016short,lee2017speeding}, represented by $\mathbb{P}\{\texttt {T} > t\} = e^{-\lambda (x - s)}$ for $ t>\varDelta$. $\lambda$ is the computation intensity parameter and $\mu = 1/\lambda$ denotes the disturbance parameter. Larger $\mu$ indicates likelihood of stragglers. The computational capacity of a node can be effectively described by the parameters $(\Delta, \lambda)$ \cite{2020Diversity}. 

Building upon the CU-based model, we analyze the time model of computing tasks, recognizing that the task's execution time is intrinsically linked to the quantity of CUs it comprises. The computational share (fraction) of $w_i$ is represented by $\gamma _i[n] =\texttt{W}_i \frac { r * \ell} {mk}$. This suggests that a $\mathrm{Job}_n$ containing $r* \ell$ CUs is divided into $m*k$ portions, with $\texttt{W}_i$ tasks delegated to the $w_i$. It's important to note that the allocation of computational shares hinges on encoding and scheduling strategies. To illustrate this, we detail two models.

Model 1: Worker-dependent scaling model. The running time of the computation task $\texttt{W}_i$ is represented by a random variable $\texttt{Y}_i$, $\texttt{Y}_i \sim \mathop {\mathrm {S-Exp}}\nolimits (\varDelta_i, \lambda_i)$.\footnote{While we've omitted the round information 'n', it's important to note that the allocation strategy does depend on 'n'.} 
$\varDelta$ denotes the minimum execution time, indicating the necessary time overhead, $\varDelta_i[n]=\varDelta_i*\gamma _i[n]$, $\varDelta\in \mathbb {R}$,  $\lambda_i[n]=\lambda_i/\gamma _i[t]$, $\lambda\in \mathbb {R}$, the Probability Distribution Function(PDF) is ${f}_{\texttt {W}_i} (\lambda_i[n],\varDelta_i[n])$.

Model 2: Task-dependent additive scaling model. This model views the entire task's computation as an interdependent process of computing CUs. Thus, the computation time $\texttt {X}_{i} = \sum _{j=1}^{\texttt {W}_i} \texttt {T}_{j}$, that is, the calculation time $\texttt {X}_{i}$ obeys the Gamma distribution.

However, it is pertinent to recognize that these models primarily facilitate analysis, and they may not accurately capture the running time dynamics of EN workers. Consequently, Section IV introduces a scheduling methodology that does not rely on any specific running time model.

\begin{figure}[!t]
	\centering
	\includegraphics[width=3.5in]{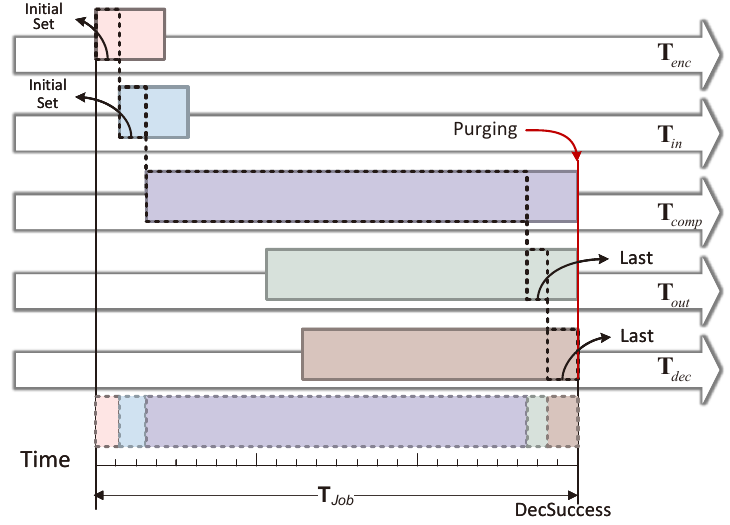}
	\caption{Job execution time.}
	\label{ELECS_fig04}
\end{figure}

\subsection{Job Execution Time}
This subsection delineates an analysis of job execution time, which is segmented into five distinct periods and grouped into three categories: encoding/decoding time ($T_{enc}$/$T_{dec}$), IN/OUT transmission time ($T_{in}$/$T_{out}$), and worker computing time ($T_{comp}$).

Leveraging the flexibility of the Rateless Encoder, an arbitrary number of encoding packets can be generated, thereby there is no need to wait for the entire job's encoding to be completed. We adopt a Quick Launch Strategy (QLS) where an initial set of CUs is encoded and distributed to the worker for computation. This procedure effectively minimizes the workers' idle time, an inefficiency commonly associated with encoding and transmission delays. The Gantt chart depicted in Figure.\ref{ELECS_fig04} illustrates the efficiency gain, particularly apparent within the overlapping segments of $T_{enc}$, $T_{in}$, and $T_{comp}$. Computation is performed with the CU as the granularity, allowing the worker to perform computations and return the results without delay. Considering the decoder requires $N$ computational results to finalize decoding, the moment of job completion is marked by the final CU's decoding. Upon successful decoding, the controller issues a purging command, initiating the clearing of $\mathrm{Job}_n$ tasks from the worker cache queue. This action readies the system for the subsequent computation cycle of $\mathrm{Job}_{n+1}$, therefore, the duration of these time periods is not simply a sum of individual phases.

To minimize signaling overhead, a strategy devoid of purging may be selected. With this strategy, worker nodes persist in processing queued tasks until their completion. As such, the computation completion time is contingent upon the volume of tasks distributed among the different nodes.

Section III provides a comprehensive analysis of the system model, dissecting crucial aspects such as workload, codec, worker, and job execution time. This discussion lays the groundwork for our exposition of the REDC processes and strategies in Section IV, which delves into the specifics of offloading workloads to edge networks with the objective of improving computational efficiency.

\section{Adaptive Job-Encoded Distributed Offloading}
This section focuses on the strategy parameters of job encoding offloading. The main goal is to optimize the workload throughput of the edge network and utilize the computing power of the working cluster to minimize the sequential execution delay in iterative jobs. In the face of inherent instability of network conditions and node states within edge networks, our strategy provides a flexible approach to task offloading. Additionally, our comprehensive strategy is fully compatible with rateless encoders, consequently augmenting the system's overall efficiency.

\subsection{Edge network worker selection and workload restrictions}
In edge networks, workers exhibit heterogeneous and dynamic characteristics. Not every available worker can enhance the overall performance of distributed computing; indiscriminate selection for the worker can lead to energy wastage and inefficient use of computational resources. Therefore, prior to offloading subtasks, it is essential to select a high-performance subset of workers $\{\mathcal I\}$, from the total set of $N_I$ workers.

We adopt the inter-arrival time, represented as $\frac{1}{\upsilon}$ time slots, as our unit to examine the state transitions during each job execution phase, focusing specifically on the transmission, codec, and computation periods.

First, we examine the transmission characteristics of each node $w_i$ for $i \in \{1, {\dots } N_I\}$. We assume a constant data transmission volume per unit time, denoted as ${b_i}$, which stands for both the uplink and downlink bandwidth of node $i$. Given the initial uplink task volume ${D_{\mathrm{in}}}$ and the final downlink task volume ${D_{\mathrm{out}}}$ for each job, we define the corresponding transmission periods as ${T_{\mathrm{in}}} = \frac{\upsilon{D_{{\mathrm{in}}}} }{{b_i}}$ and ${T_{\mathrm{out} }} = \frac{{\upsilon{D_{{\mathrm{out}}}} }}{{b_i}}$, respectively.

Subsequently, we investigate the computational capabilities of the encoding and decoding processes, represented as ${p_{\mathrm{enc}}}$ and ${p_{\mathrm{dec}}}$ respectively. The computational workload required for these processes is denoted by ${C_{\mathrm{enc}}}$ and ${C_{\mathrm{dec}}}$ respectively. Hence, the average computation durations for encoding and decoding are defined as ${T_{\mathrm{enc}}} = \mathbb{E}\left[\frac{{C_{\mathrm{enc}}}}{p_{\mathrm{enc}}}\right]$ and ${T_{\mathrm{dec}}} =\mathbb{E}\left[\frac{{C_{\mathrm{dec}}}}{p_{\mathrm{dec}}}\right]$.

Finally, we analyze the computational characteristics of each worker $w_i$. The time taken by $w_i$ to process a job is denoted as $T_i$, a random variable that may adhere to a variety of distributions including but not limited to offset exponential, Gamma, Pareto, and Bi-distributions. We denote the average processing time for node $w_i$ as $T_{wi}$, thus $T_{wi} = \mathbb{E}[T_i]$.

In the processing of workloads, we identify computation abilities as the potential bottleneck. If overheads other than computation become too large, they will diminish the efficiency of offloading tasks, increase energy consumption, and result in wastage of edge computational resources. Therefore, we focus on nodes whose transmission capabilities supersede their computational abilities. In addition, the encoders and decoders utilized in this study are designed to have linear computational complexity, which should be significantly lower than the computational complexity inherent to the tasks themselves. Based on these considerations, we define the discriminant formula for valid workers as follows:
\begin{equation} \label{ValidWorker}
	\min \left ( { \frac {1}{ T_{\mathrm{enc}}},\;\frac {1}{ T_{\mathrm{dec}}},\;\frac {1}{ T_{\mathrm{in}}},\;\frac {1}{ T_{\mathrm{out}}} }\right)\geq \frac {1}{ T_{wi}},
\end{equation}
%among them, $\mathbb E\left[ r^{\mathrm{enc}} \right]$ is the average rate for the encoder to generate $\mathcal N$ encoded symbols, and ${ \mathbb E\left [{r^{\mathrm {dec}}}\right]}$ is the average rate for the decoder to recover $K$ source symbols.

To prevent congestion in the edge network,let $\Gamma = (1+\epsilon)$ denote the inverse of the code-rate, the arrival rate $\Gamma \upsilon$ of encoded jobs carrying workloads should not exceed the overall processing rate of the edge network

\begin{equation} \label{ENStability}
	\Gamma \upsilon \leq \sum _{i=1}^{N_I} \frac {1}{T_{wi}}\cdot
\end{equation}

Under a specified distribution policy $\gamma_i$, where $i \in {1,\ldots,\mathcal{I}} $, the job arrival rate at each node $w_i$ is $\upsilon \gamma_i$. To guarantee the stability of each worker and prevent buffer queue overflow, this rate should not exceed the processing capacity of the respective node.
\begin{equation} \label{WorkerStability}
	\upsilon \gamma _{i} \leq \frac {1}{T_{wi}},\quad \forall i\in \{1,\ldots,\mathcal{I}\}.
\end{equation}

We introduce $r_{i}^{\text {comp}} = \frac{1}{\upsilon T_{wi}}$, which sets a limit on the proportion of workload assigned to each worker node, expressed as $\gamma_{i} \leq r_{i}^{\text {comp}}$, $\forall i\in \{1,\ldots,\mathcal{I}\}$. With $\sum _{i=1}^{\mathcal{I}}\gamma _{i}=\Gamma$, equation \ref{WorkerStability} assures the satisfaction of equation \ref{ENStability}, thereby guaranteeing the overall system stability. These form the fundamental constraints of our strategy.

\subsection{Workload Offloading Strategy}
Under the scheduling strategy $\gamma_{i}$, each worker is assigned a proportionate share of the job. We utilize $C_{\mathrm{Job}}$ and $C_{\mathrm{Task}}$ to represent computational complexity, such that $C_{\mathrm{Task}} = \gamma_{i} C_{\mathrm{Job}}$, the master node can estimate the real-time computational power status information based on feedback from the workers, specifically $\mathbb E [T_{i}]$ and $\mathbb E[T_{i}^2]$. 

In the heterogeneous edge network, a controller manages the task of distributing the computational load across varied workers. Each $i$th worker receives a proportion, $\gamma_i \in [\underline {\gamma }_{i},1]$, $\forall i\in \{1,\ldots,\mathcal{I}\}$, of the total workload such that $\sum_{i = 1}^{\mathcal {I}} \gamma_i = \Gamma$, $\Gamma \in [1,N_I]$.
To formulate an efficient and concise scheduling strategy, the controller employs an M/G/1 queuing model for modeling and analyzing the workers. Each worker node has a job arrival rate of $\upsilon \gamma_i$, with $\mathbb E[T_i]$ and $\mathbb E[T_i^2]$ representing the first-order moment and second-order moments of its service time, respectively. The workload on the $i$th worker queue is given by $\rho_i = \upsilon{\gamma_i \mathbb E [T_i]}$.

Consequently, the average response time for the $i$th worker to process a task, encompassing both the queuing wait time and the actual processing time, is determined by the P-K theorem\cite{pollaczek1930uber,cohen2021stream}.

\begin{equation}
	\begin{split} 
		L_{\text {comp},i} =&\frac{\rho + \upsilon \mu \mathrm{Var}(S)}{2(\mu - \upsilon)}+\mu ^{-1}\\
		=&  \frac {\upsilon \gamma _{i} {\mathbb E}\left [{T_{i}^{2}}\right]}{2 \left ({1-\rho _{i}}\right)}+ {\mathbb E} \left [{T_{i}}\right]\\=&\frac {1}{\upsilon } \left ({\frac {a_{i}\gamma _{i}}{ r_{i}^{\text {comp}}-\gamma _{i}}+\frac {1}{ r_{i}^{\text {comp}}} }\right).
	\end{split}
\end{equation}

Let's note that $a_{i} = \upsilon {\mathbb E[T_{i}^{2}]}/{2\mathbb E[T_{i}]}$. Utilizing this, we can calculate the average Job computational delay across the entire edge network. As all workers perform computations simultaneously, the average delay across all workers defines the overall Job computation delay.

\begin{equation} 
	L_{\text {comp}}=\frac {1}{\mathcal {I}}\sum _{i=1}^{\mathcal {I}} \frac {1}{\upsilon } \left ({\frac {a_{i}\gamma _{i}^{2}}{ r_{i}^{\text {comp}}-\gamma _{i}}+\frac {\gamma _{i}}{ r_{i}^{\text {comp}}} }\right).
\end{equation}

In addition, we define the communication capability, $ r_{i}^{\text {comm}} = b_{i}/\upsilon(D_{\text {in}}+D_{\text {out}})$, 
\begin{equation} 
	L_{ \text {comm}} = \frac {1}{\mathcal {I}}\sum _{i=1}^{\mathcal {I}}\frac {\gamma _{i}}{\upsilon r_{i}^{\text {comm}} },
\end{equation}

This represents the transmission rate of each node, including both uplink and downlink data transmission.

So the average calculation time overhead of Job is
\begin{equation} \label{ExeTime}
	L_{\text {exe}} =L_{ \text {comp}}+L_{ \text {comm}}+\mathbb E\left [{T_{\text {enc}}}\right]+ {\mathbb E\left [{T_{\text {dec}}}\right].}
\end{equation}
%Ultimately, the total execution time of a job can be encapsulated as:$T_{\mathrm{Job}} = \text{avg} (T_{in}+ T_{out}+T_{enc}+T_{dec})+ \text{max} { T_{\mathrm{comp}}}$.

Given the dynamic heterogeneity of edge network workers, defining an optimal allocation strategy that effectively balances the workload among them is crucial. Such a strategy seeks to minimize the average job execution time and maximize computing efficiency. The proposed strategy is:

\begin{equation} \label{OptimalSplit}
\begin{split} 
	\boldsymbol {\gamma }^{\star } = \underset {\boldsymbol {\gamma }}{\arg \min }&\sum _{i=1}^{\mathcal{I}} \left ({\frac {a_{i}\gamma _{i}^{2}}{ r_{i}^{\text {comp}}-\gamma _{i}}+\left ({\frac {1}{ r_{i}^{\text {comp}}}+\frac {1}{ r_{i}^{\text {comm}}}}\right)\gamma _{i} }\right)\Gamma,\\ \text {s.t.}&\sum _{i=1}^{\mathcal{I}}\gamma _{i}=\Gamma,\\&\underline {\gamma }_{i}\leq \gamma _{i}, \qquad \forall i\in \{1,\cdots,{\mathcal{I}}\}, \\&\gamma _{i} \leq r_{i}^{\text {comp}},\quad \forall i\in \{1,\cdots,{\mathcal{I}}\},
\end{split}
\end{equation}

The optimization goal, aimed at minimizing the overall job execution time, is subject to the following constraints: 1) $\gamma$ should represent a partition of the total workload, i.e., positive values that sum to $\Gamma$;
2) The minimum task allocation ratio, or the use of $\{\mathcal I\}$ workers, must reach a non-negative value denoted as $\underline {\gamma }_{i}$;
3) Each worker must maintain a stable queue to prevent tasks from arriving faster than they can be processed, avoiding buffer overflow.

The solution to this optimization problem - the allocation strategy, denoted as $\gamma _{i}^{*}$, can be formulated as:

\begin{equation} 
\begin{split} \gamma _{i}^{*}=\!\!\begin{cases} \max \left \{\Gamma { r_{i}^{\text {comp}}\left ({1-\sqrt {\frac {a_{i}}{a_{i}+\eta -\frac {1} { r_{i}^{\text {comp}}}-\frac {1}{ r_{i}^{\text {comm}}}}}}\right), \underline {\gamma }_{i}}\right \},\\ \qquad \qquad \qquad \qquad \qquad \frac {1}{ r_{i}^{\text {comp}}}+\frac {1}{ r_{i}^{\text {comm}}}-a_{i} < \eta,\\ \underline {\gamma }_{i},\qquad \qquad \qquad \qquad ~~\text {otherwise.} \end{cases}
\end{split}
\end{equation}
where $\eta$ is set such that $\sum _{i=1}^{\mathcal I} \gamma _{i} = \Gamma$.
\subsection{Coding Strategy}
This part examines the Distributed Computing Rateless Encoder (RE). The strategy can be depicted as RE($K, \Omega(x), \mathcal {N}$), where $\mathcal {N}=\Gamma K$, indicates the number of encoded symbols output by the encoder. We first analyze the task redundancy ratio $\Gamma$, influenced by both decoding and stragglers-resilience overhead. The controller's regulation of encoding symbol quantity directly affects system performance. Below we define and discuss these two overheads in detail.

Decoding overhead is denoted as $\epsilon_{\mathrm{dec}}$, such that $N=K(1+\epsilon_{\mathrm{dec}})$. Influenced by three factors: decoding success probability $y_l$, source symbol count $K$, and degree distribution $\Omega(x)$.  Differing from decoding methods such as MDS\cite{lee2017speeding}, Matdot\cite{2018On}, where a specific number of coded symbols assures decoding, the fountain code's characteristic implies that received coded symbols only confer a probability of source recovery. The probability of successful recovery increases with the number of coded symbols received. Consequently, given a decoding success probability,  $\epsilon_{\mathrm{dec}}$ can be determined via density evolution, provided $K$ and  $\Omega(x)$ are fixed. The process of density evolution analysis is detailed in the appendix A. Typically, when $K > 10000$, $\epsilon_{\mathrm{dec}}$ falls below 0.011.

Stragglers-resilience overhead is denoted as $\epsilon_{\mathrm{\mathcal S}}$, plays a significant role in performing distributed computing in edge networks with stragglers, a scenario comparable to data transmission in erasure channels. Here, redundant data becomes vital to ensure resilience against stragglers. Importantly, this issue requires careful balancing: Leveraging the prefix feature of fountain codes, a larger $\epsilon_{\mathrm{\mathcal S}}$ indicates stronger resistance against interference, but also augments the system's burden.

In light of the above, we can derive $\Gamma = 1+\epsilon =1+ \epsilon_{\mathrm{dec}} + \epsilon_{\mathrm{\mathcal S}}$. The encoder's output quantity $\mathcal N$ can be determined by assessing decoding success probability and stragglers-resilience. Typically, encoding done in rounds, with data volume defined by values $\aleph$ and $\mathcal T$ in Algorithm \ref{alg:algorithm1}. It's noteworthy that, without taking into account the interaction overhead, the encoder's adaptive output based on responses could, in theory, maximize stragglers-resilience.

\begin{algorithm}[!t]
	\caption{Rateless Encoding Distributed Computing}
	\label{alg:algorithm2}
	\KwOut{Optimal REDC Strategy $\mathbf{\mathcal S}$ include:	RE$(K,\mathbf{\Omega}(x), \mathcal N)$, $\mathcal I^*$ and ${\{\gamma_1,...,\gamma_{\mathcal {I}}\}} ^ *$}
	$\textbf{Initialization}$: $L_{exe}^*$ = $\infty$\;
	\For{$\mathrm{RE}\{(K,\mathbf{\Omega}(x), \mathcal N)\} \in  \mathbf{\mathcal S}$}
	{	
	Determine \(\mathcal I\) using the condition specified in Eq. \ref{ValidWorker}\;
	Find the optimal split \(\{\gamma_1,...,\gamma_{\mathcal I}\}^*\) using Eq. \ref{OptimalSplit}\;
	Compute the execution time \(L_{exe}\) using Eq. \ref{ExeTime}\;
		\If{$L_{exe} < {L_{exe}^*}$}
		{
			{DCRE$(K,\mathbf{\Omega}(x),\mathcal N)^*$} = {DCRE$(K,\mathbf{\Omega}(x),\mathcal N)$}\;
			${\{\gamma_1,...,\gamma_{\mathcal {I}}\}} ^ *$ = ${\{\gamma_1,...,\gamma_{\mathcal {I}}\}}$\;
			$\mathcal I^*$ = $\mathcal I$\;
		}
	}
\end{algorithm}

\subsection{Determine REDC Strategy Parameters}
Given the offloading strategy $\mathbf{\mathcal S}$ utilized by REDC in the edge network, the process for determining its essential parameters is demonstrated in Algorithm 2. The first step involves choosing a suitable group of workers in the edge network. This entire selection maintains a redundancy of $\varPhi$ to ensure support for the decoding redundancy of $K \epsilon_{dec}$. Based on the status feedback, we sort accessible workers according to their computing power $r_{i}^{\text {comp}}$, and the resultant set is expressed as $\bar{I}$. We select the first $\mathcal I$ nodes as the optimal computing node set ${\mathcal I}$, ensuring that the chosen set meets the computing power requirements.

\begin{equation} 
	\sum _{i=1}^{\mathcal I} r_{i}^{\text {comp}}\geq 1+ \varPhi \quad \text {s.t.}\quad \varPhi \geq \epsilon_{\mathrm{dec}}.
\end{equation} 

In our model, given the job arrival rate of $\upsilon$, the computational power of the edge network is deemed sufficient, making the aforementioned selection requirements feasible. The $r_{i}^{\text {comp}}$ is contingent on the encoding parameters. We represent the set of encoding strategies with {RE$(K,\mathbf{\Omega}(x), \mathcal N)$}$\in \mathbf{\mathcal S}$, where each parameter alteration in $\mathcal S$ corresponds to a distinct strategy.

Indeed, the heart of the problem lies in determining the encoding parameters. Given a set of encoder strategies, we can ascertain parameters such as $D_{\text{in}}$, $D_{\text{out}}$, $C_{\text{job}}$, ${C_{\mathrm{enc}}}$ and ${C_{\mathrm{dec}}}$. Using these values, we can calculate $L_{\text{exe}}$ Eq.\ref{ExeTime}. The strategy associated with the shortest execution time is logged as $S^{*}$. This process only needs to be run when the access state changes, and given the confined solution space,  an exhaustive search algorithm is employed to find the optimal results. By retaining the information of historical policies, policies can be directly established for the same state.

Because a change in encoding strategy impacts $C_{\text{job}}$, $E(T_i)$ will also alter, even with the same scheduling strategy. Hence, information about $T_i$ and $T_i^2$ needs to be maintained through state feedback. This necessitates the execution of a few jobs in advance upon initial workload reception to obtain these state information. For specific state information maintenance methods, refer to Section IV-E. 

The computation time of Algorithm \ref{alg:algorithm2} needs to be substantially shorter than the job arrival rate $\upsilon$ to ensure the algorithm's overhead is acceptable. Moreover, due to the flexibility of the fountain code, each node can quickly complete the estimation of node state information by performing granular feedback time calculations through CU. This significantly reduces the computational overhead of Algorithm 2.

We hereby recall the fundamental elements of our workload. Each job is a matrix multiplication, ${{\mathbf{A}}^T} \times {\mathbf{B}}$, where $A\in \mathbb{F}^{s \times r}$, and $B\in \mathbb{F}^{s \times \ell}$. The quantity of source symbols implicated in the encoding process is represented as $K=m*k$. Furthermore, the number of encoding symbols required for the decoding process is symbolized by $N$, expressed as $N=K(1+ \epsilon_{\mathrm{dec}})$. Here, $\epsilon_{\mathrm{dec}}$ depends on our expected decoding success rate,  The methodology to calculate this value is elucidated in Appendix A.

Upon revisiting the Quick Launch Strategy outlined in Section III-E, it's important to underscore that this strategy expedites the commencement of computations by encoding and transferring only an  initially subset of computation units. 
Moreover, we opt for nodes with superior computational rate \ref{ValidWorker} while thoroughly taking into account both the codec and transmission rates. Under the REDC framework, once a task begins execution, it proceeds uninterrupted. Throughout the job execution cycle, the scheduler ensures a non-empty node queue, thus obviating any need for task pausing. Upon completion of each CU's computation, the result is promptly forwarded back. Consequently, our swift startup solution restricts the transmission time overhead to the data volume of an individual computation task\footnote {The encoding vector is not considered}. Accordingly, $D_{\mathrm{in}} =\frac{rs}{m} +\frac{\ell s}{k}$, which corresponds to the cumulative size of a single encoding symbol, given by $\mathbf {A}_{m} \in {\mathbb{F}}_{s \times r/m}$ and $\mathbf {B}_{k} \in {\mathbb{F}}_{s \times \ell/k}$.
Moreover, $D_{\mathrm{out}} = \frac{r\ell}{mk} $ signifies the size of a single operation result, represented by $ \mathbf {\bar C} \in {\mathbb{F}}_{r \times \ell }$.

In a similar vein, the encoding complexity of the Quick Launch Strategy only necessitates consideration a subset of $\{\mathcal I\}$ 's encoding. The encoding operation is the multiplication of the coefficient vector by the source symbol, the number of addition operations being equivalent to the weight of the coefficient vector. Given that the encoding vector weights $d'$ and $d''$ of $\{\mathbf {A}_{m}\}$ and $\{\mathbf {A}_{k}\}$ are not directly expressible by the degree distribution, obtaining a definite value for ${C_{\mathrm{enc}}} =\mathcal I *(\frac{d'rs}{m} +\frac{d''\ell s}{k})$ poses a challenge. However, considering the average degree of the output symbol, denoted as $\beta = \Omega' (1)$, we can ascertain the upper and lower bounds of $C_{enc}$ as $\mathcal I \beta (I_{in})$ and $\mathcal I \ \sqrt{\beta} (I_{in})$, respectively.
${C_{\mathrm{enc}}}$ and ${C_{\mathrm{dec}}}$
The decoding computation complexity is evaluated as ${C_{\mathrm{dec}}} =D_{\mathrm{out}} * K \ln K$. We can compute the complexity after receiving last encoded symbol, a single decoding process is the normalization of the number of symbols $N$, as $\frac{D_{\mathrm{out}} K \ln K}{N}$, resulting in $\frac{D_{\mathrm{out}} \ln K }{1+\epsilon_{\mathrm{dec}}}$.

\subsection{Adaptive Estimation of Workers’ Computational Statistical Features}
Given the fluctuating state of edge network computing nodes, it's imperative to adjust load distribution adaptively based on node status. Thus, the controller should generate scheduling strategies base on node status information $\mathbb E[T_i]$ and $\mathbb E[T_i^2]$.

A central component of our proposed approach is the maintenance of node state information, specifically  $\mathbb E[T_i]$ and $\mathbb E[T_i^2]$ for each node $i$ in the set $\{1,…,N_I\}$. This state information forms the foundation for the controller's scheduling policies. By tagging the start and end times of tasks, and reporting these along with the computational results and state feedback, we are able to ascertain the execution duration task completion, thereby allowing us to estimate the current node's computational capacity. 
We denote worker nodes' feedback computational time, reflecting task execution duration, as ${\varpi_{i}}$, enabling the master node to dynamically update state information. 
We apply the Exponentially Weighted Moving Average (EWMA) method to upkeep state information, symbolized as $M_{i}(t)$ and $V_{i}(t)$. These represent the first and second moment estimates of ${{\varpi_{i}}}$ at time $t$, respectively, computed as such:
\begin{equation} \label{EWMA}
\begin{split}
	M_{i}(t) & =\alpha M_{i}(t - 1) + (1 - \alpha) {\varpi_{i}}(t), \\
	V_{i}(t) & = \beta V_{i}(t - 1) + (1 - \beta){\varpi_{i}}^{2}(t).
\end{split}
\end{equation}

\begin{figure*}[h]
	\centering
	\subfloat[$K=49$.\label{result1:sub1}]{
		\includegraphics[width=0.31\textwidth]{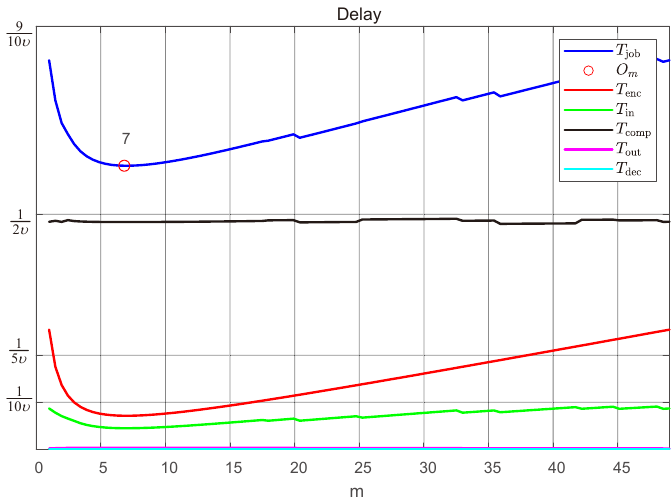}}
	\hfill
	\subfloat[$K=100$.\label{result1:sub2}]{
		\includegraphics[width=0.31\textwidth]{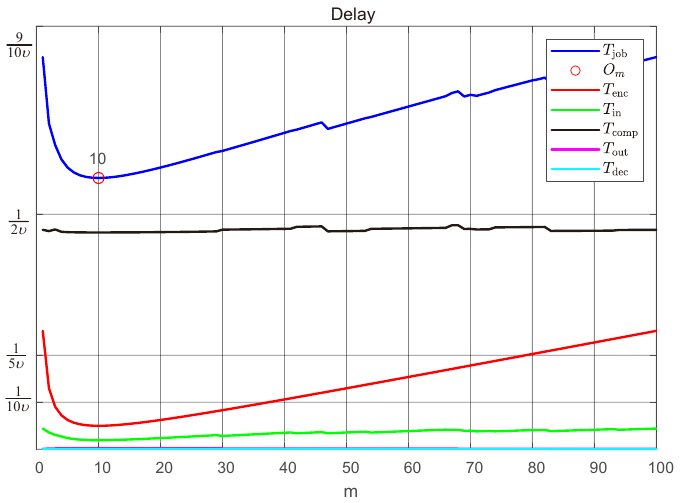}}
	\hfill
	\subfloat[$K=225$.\label{result1:sub3}]{
		\includegraphics[width=0.31\textwidth]{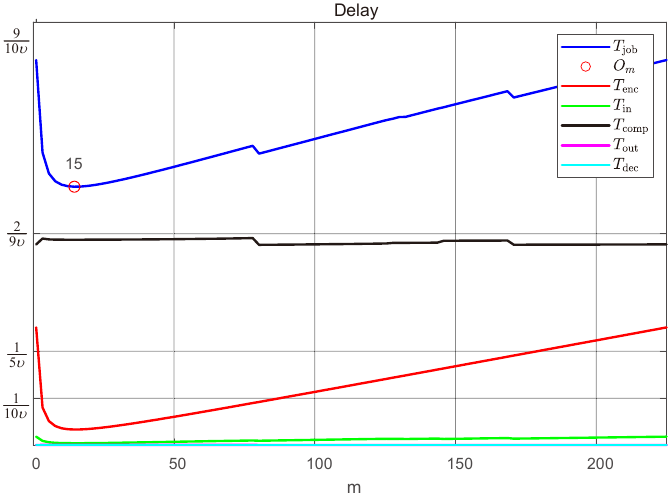}}
	\caption{Each subplot represents the variation of $T_{\mathrm{job}}$ with respect to the parameter $m$ for different strategy parameters ($K = mk$). $O_m$ indicates the $m$ value minimizing $T_{\mathrm{job}}$.}
	\label{result1}
\end{figure*}

\section{Performance Results}
This section examines the REDC method through numerical analysis, focusing on the encoding strategy, the QLS, adaptive encoding performance, overall resource utilization, and adaptability. The method is compared with a previous scheme cited in \cite{cohen2021stream}.
The simulations were conducted in a MATLAB environment running on a desktop computer, equipped with a Core I7-10700 processor and 32GB of RAM. We examined matrix multiplication calculations using square matrices of dimensions $r,s,\ell$=100 for representative purposes without loss of generality. In the simulated edge network, we assumed the presence of 150 accessible nodes with a maximum selection cap of 100 operational nodes per instance. The task-splitting approach employed was based on the strategy $K=mk$.

For these simulations, the service time for each node was derived from one or more distribution functions. In scenarios devoid of straggler characteristics, the task service time adheres to an exponential distribution. Recognizing the inherently unpredictable nature of the edge network node statuses, we introduced characteristics from the Pareto and bi-distributions. The characteristics of these distributions are illustrated in Figure.\ref{ELECS_fig02}.b.

We implemented Model 2 as outlined in Section III-B. This model takes into consideration the quantity of CUs assigned to tasks and assumes that task service times follow a Gamma distribution. Lastly, the choice of degree distribution was guided by the distribution values delineated in \cite{2020Factored}, consider a modified version of Soliton degree distribution given by $\Omega (x) = \sum_d {\Omega _d}{x^d}$, where${\Omega _d} = \frac{\omega _d}{\sum _{d=0}^{K} {\omega _d}}$,

\begin{equation}
	{\omega _d}=\begin{cases} \frac {1}{K}, d=1;\\ \frac {1}{2}, d=2;\\ \frac {1}{d(d-1)}, 3 \geq d \geq \max (m,k);\\ \frac {1}{d(d-1)}, \max (m,k) < d \leq mk, d~\text {is not prime};\\ 0, \max (m,k) < d \leq mk, d~\text {is prime}. \end{cases}
\end{equation}

\begin{figure}[h]
	\centering
	\includegraphics[width=3.5in]{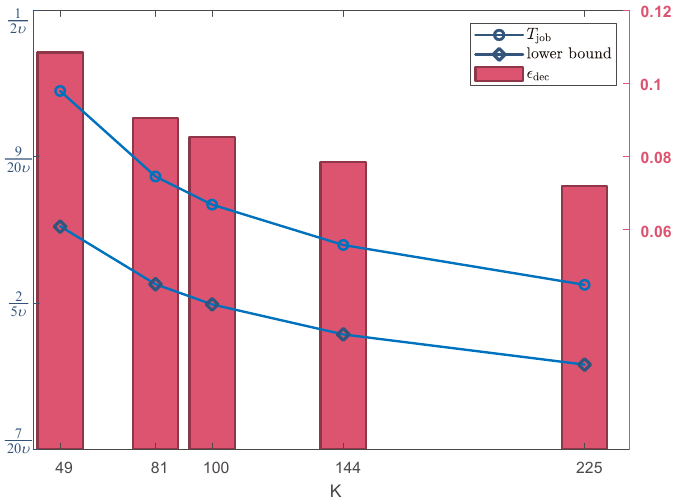}
	\caption{Relationship Between Decoding Efficiency $\epsilon_{\mathrm{dec}}$ and Job Splitting Quantity $K$.}
	\label{result2}
\end{figure}

We investigated the characteristics of the job execution life cycle, \( T_{\mathrm{job}} \) (detailed in Section IV-D), and found it to be closely related to the encoding strategy. With fixed EN nodes parameters and encoding strategies at \( K=[49,100,225] \), and \( \Gamma=[1.36,1.30,1.20] \), we randomly selected computational and bandwidth characteristic values \( \mu_i \) and \( b_i \) within [0,2500] and [0,1000], respectively. The values for \( {p_{\mathrm{enc}}} \) and \( {p_{\mathrm{dec}}} \) were set at 10000 and 1000. Our results (Figure.\ref{result1}) demonstrate that the minimal value $O_m$ of \( T_{\mathrm{job}} \) is obtained when \( m=\sqrt K \).
Furthermore, we evaluated our Quick Launch Strategy (QLS) by comparing it to a conventional fully-encoded-before-distribution method. As shown in Figure.\ref{ELECS_fig04}, for \( K=[49,100,225] \) and \( m=\sqrt K \), QLS required only \( [88.94\%, 85.56\%, 83.24\%] \) of the time, thereby capitalizing on the flexibility of rateless codes.

We selected $m=k=\sqrt K$ for our analysis and consider the purging mode, which entails clearing the task queue after job decoding is complete. Next, we examined the time required for actual decoding to reach completion. As illustrated in the line plot of Figure.\ref{result2}, and as detailed in line 12 of Algorithm \ref{alg:algorithm1}, the coefficient matrix $\mathbf{Z}$, obtained by the decoder via the operation $\mathbf{Z} \leftarrow \mathbf{append}[\mathbf{Z},\vec{z}]$, becomes column is full rank was recorded. For five splitting values of $K=49-225$, we conducted 500 distributed Job operations and obtained the system's average running time $T_{\text{job}}$. It is noteworthy that the most ideal lower bound of the system is $N=K$, signifying that distributed computing does not require redundancy at this juncture.  A histogram was utilized to display the redundancy of our method, revealing a decrease in redundancy with increasing $K$, in alignment with our expectations. The average value of $\epsilon_{\text{dec}}$ was determined to be 0.087.

\begin{figure*}[h]
	\centering
	\subfloat[Worker Contribution Percent.\label{result3:sub1}]{
		\includegraphics[width=0.31\textwidth]{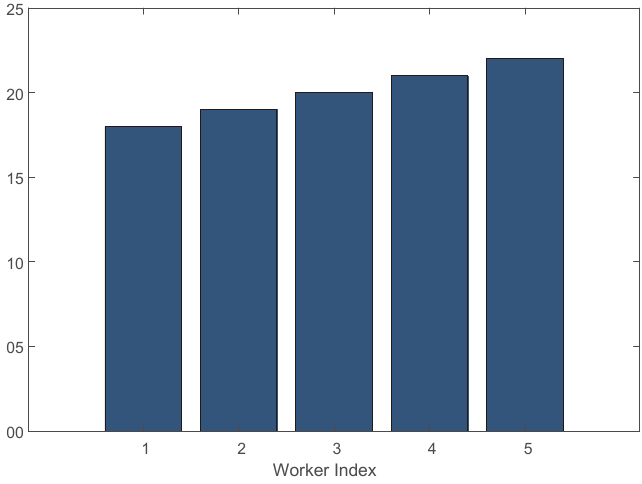}}
	\hfill
	\subfloat[$r_{\mathcal I}^{\text {comp}} \{\varPhi = 0.09\}$.\label{result3:sub2}]{
		\includegraphics[width=0.31\textwidth]{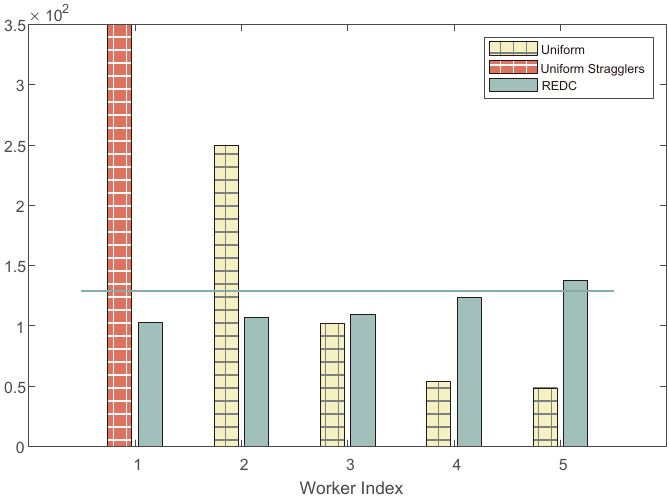}}
	\hfill
	\subfloat[$r_{\mathcal I}^{\text {comp}} \{\varPhi = 0.50\}$.\label{result3:sub3}]{
		\includegraphics[width=0.31\textwidth]{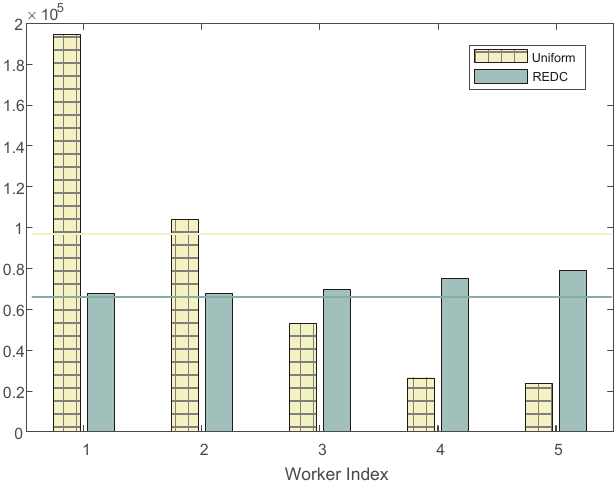}}
	\caption{Comparative analysis of REDC scheduling and uniform distribution within EN distributed computation across heterogeneous nodes.}
	\label{result3}
\end{figure*}

\begin{figure*}[h]
	\centering
	\subfloat[Worker Contribution Percent.\label{result4:sub1}]{
		\includegraphics[width=0.31\textwidth]{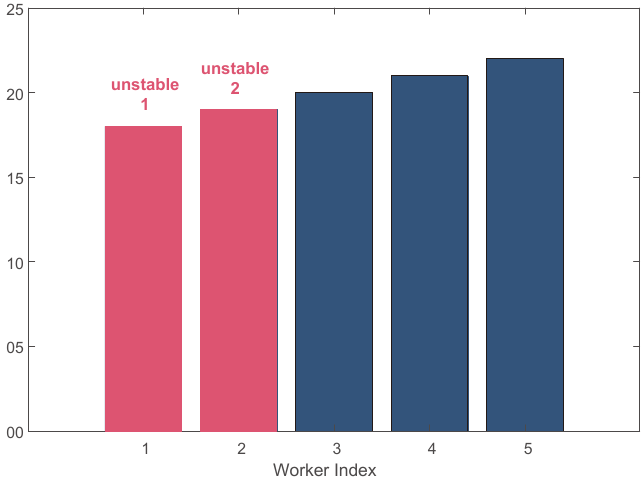}}
	\hfill
	\subfloat[$r_{\mathcal I}^{\text {comp}} \{\varPhi = 0.09\}$.\label{result4:sub2}]{
		\includegraphics[width=0.31\textwidth]{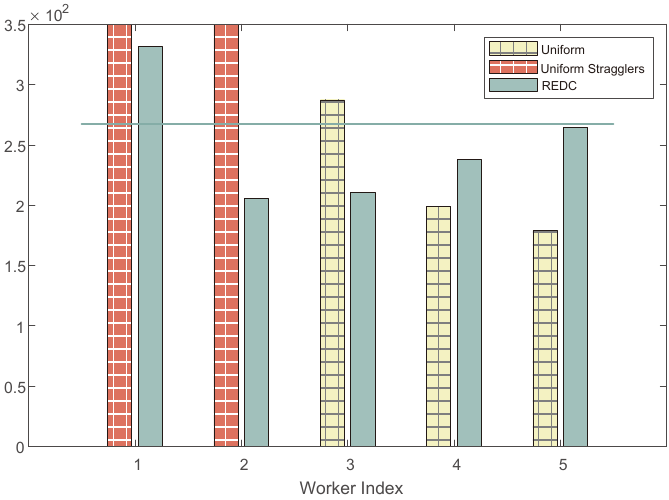}}
	\hfill
	\subfloat[$r_{\mathcal I}^{\text {comp}} \{\varPhi = 0.50\}$.\label{result4:sub3}]{
		\includegraphics[width=0.31\textwidth]{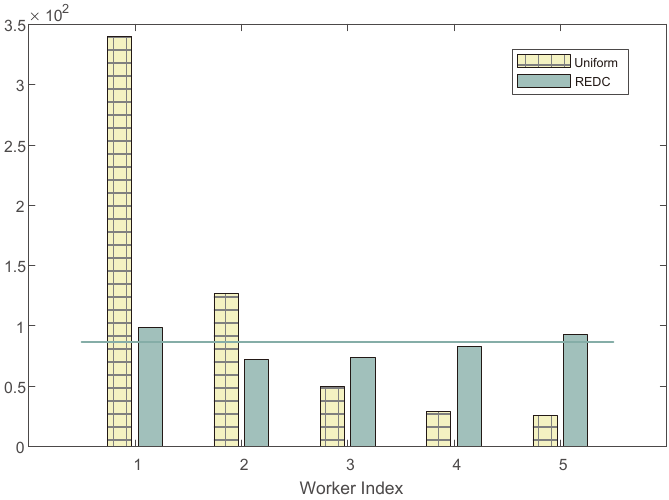}}
	\caption{Comparative analysis of REDC scheduling and uniform distribution within EN distributed computation across unstable nodes.}
	\label{result4}
\end{figure*}

Figure.\ref{result3} evaluates REDC scheduling and coding efficiency with \(\mathcal I\) nodes, given \(5\) valid nodes \(\mathcal I\) as per Eq.\ref{ValidWorker}. A fixed computational power contribution ratio is applied, where the total contribution equals \(1\), as shown in Figure.\ref{result3:sub1}. To represent varying node characteristics, two \(\varPhi\) values were compared for distributed computation with uniform distribution and REDC scheduling figure.\ref{result3:sub2} \ref{result3:sub3}. For \(\varPhi = 0.09\), computational load \(\Gamma = 1 + \varPhi\) approaches the average overhead $\epsilon_{\text{dec}}$. Without extra supplemental tasks, completion depends on all task shares, and uniform distribution exhibits a stragglers effect, particularly in node 1, reaffirming that appropriate scheduling is indispensable, especially in low workload redundancy. Figure.\ref{result3:sub3} illustrates the scenario where \(\varPhi=0.50\) , \(\Gamma=1+\varPhi\), signifying an increase in redundancy as well as enhanced computational capacity. Upon analyzing completion rates across various nodes, it becomes evident that the redundancy accelerates the task completion even for uniformly distributed. However, this comes at the expense of the overall job completion time, particularly hampered by slower nodes. In contrast, REDC, leveraging higher computational redundancy, achieves more rapid task completion with a lower variance in time across all tasks. These findings underscore the efficacy of REDC's load balancing, particularly in environments characterized by sufficient computational resources and the presence of heterogeneous nodes.

In Figure.\ref{result4}, unstable nodes are utilized to simulate stochastic characteristics of node failures. Unstable1 is defined by a 95\% exponential and 5\% Pareto distribution, reflecting poor service capabilities with service blocking. Unstable2 consists of a 95\% exponential and 5\% binomial distribution, simulating node recovery from a broken link. As shown in Figure.\ref{result4:sub2} .\ref{result4:sub3}, REDC demonstrates balanced task completion even under unstable conditions. 

\begin{figure*}[h]
	\centering
	\subfloat[Average execution time versus computational load.]{\includegraphics[width=0.47\linewidth]{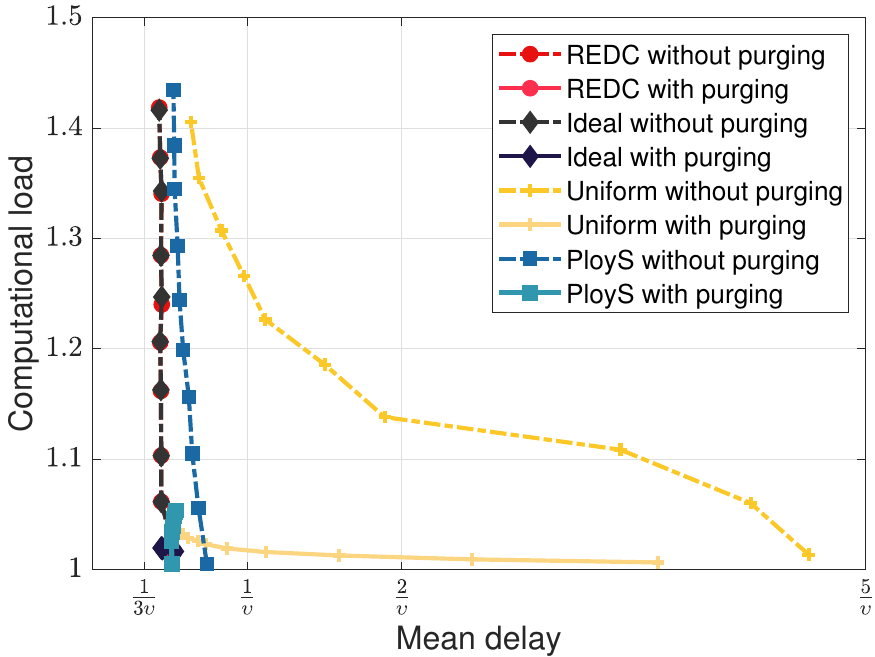}\label{result5:sub1}}\hfill
	\subfloat[Average execution time versus computational load.(zoomed-in)]{\includegraphics[width=0.47\linewidth]{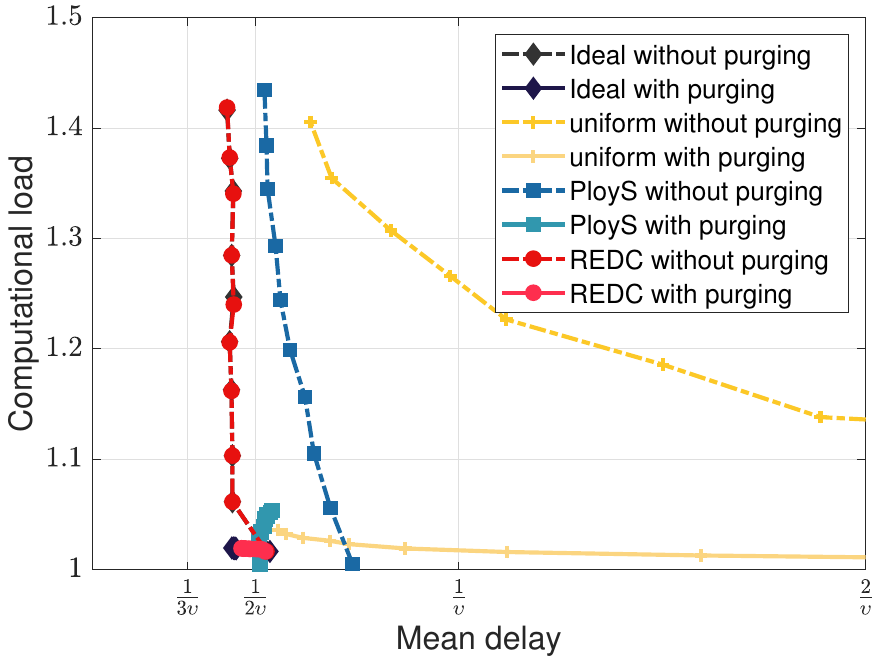}\label{result5:sub2}}\\
	\subfloat[Visualization of Execution Time in Relation to Computational Load.]{\includegraphics[width=0.47\linewidth]{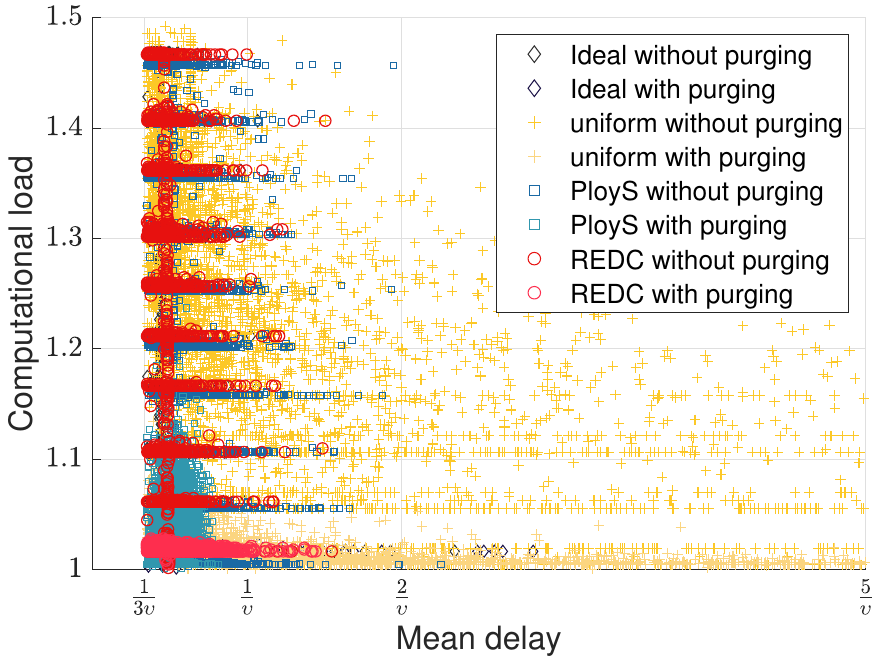}\label{result5:sub3}}\hfill
	\subfloat[Visualization of Execution Time in Relation to Computational Load.(zoomed-in)]{\includegraphics[width=0.47\linewidth]{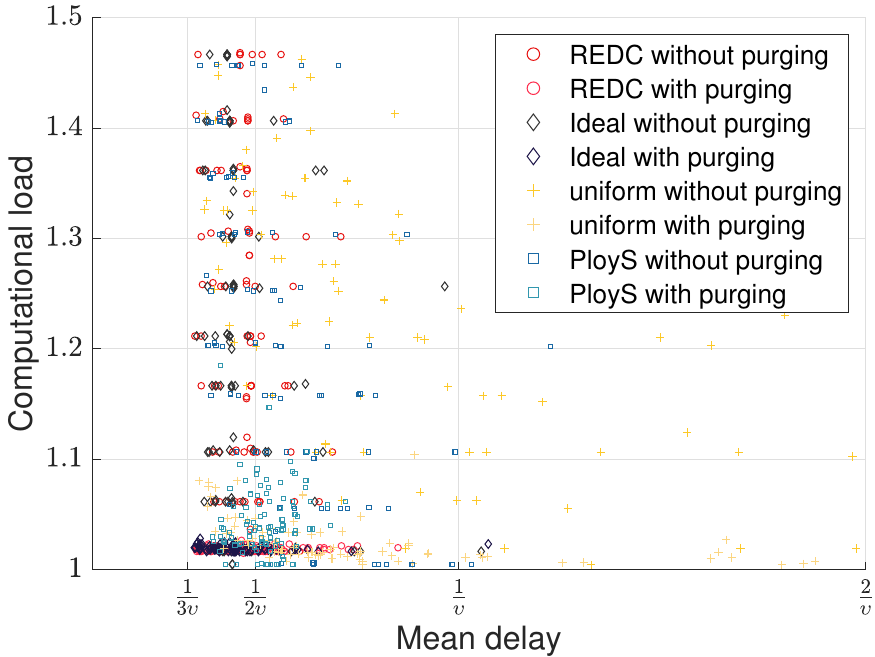}\label{result5:sub4}}
	\caption{Comparative analysis of execution time and computational load.}
	\label{result5}
\end{figure*}

In the context of evaluating the REDC, a controlled experiment was conducted encompassing three distinct strategies.  'Uniform,' where tasks are equally split among nodes; 'Ideal,' where tasks are reassigned upon completion to prevent idle nodes; and 'PolyS,' a flexible scheduling method using PolyDot code \cite{cohen2021stream}. In the simulation, computing power eigenvalue \(\mu\) and bandwidth eigenvalue \(b\) are randomly selected within \([0,2500]\) and \([0,1000]\), respectively, with \(p_{\mathrm{enc}}=10000\) and \(p_{\mathrm{dec}}=1000\). With \(\varPhi=2\) and \(\sum_{i=1}^{\mathcal I} r_{i}^{\text{comp}} = 3\), the edge network can carry \(3 \upsilon\), and a larger \(\varPhi\) increases system stability and strategy space. We define \(L_{\text{exe}}\) as the job's delivery delay, with the task arrival rate \(\epsilon = 10^{-3}\) and scheduling redundancy \(\Gamma\) in the range \([1,1.5]\). This makes the total calculation task dependent on \(\Gamma\), with \(\mathcal N=K\Gamma\). The uniform allocation scheduling coefficient is \(\gamma_i = 1/N_I\) for \(i \in {N_I}\), which varies with node selection. The ideal division ensures immediate task assignment upon node idleness, eliminating node idle states. Purging, an optional mode that enhances system flexibility and efficiency at the cost of increased signaling overhead, clears the task cache queue once decoding is complete. Figure.\ref{result5} illustrates the system execution delay and node computing load under various strategies, showing that the calculation load converges to 1 when purging is applied, due to the task quantity's influence on calculation time. 

As depicted in Fig.\ref{result5:sub2}, the average execution latency is the largest in the uniform split of node characteristics due to the instability of some nodes. Specifically, the stragglers effect becomes more pronounced when the calculation redundancy \(\Gamma\) is small. As \(\Gamma\) increases, faster nodes compensate for the calculation time, causing \(L_{\text{exe}}\) to decrease, and when \(\Gamma\) is larger, all strategies converge more closely.

\begin{figure}[h]
	\centering
	\includegraphics[width=3.5in]{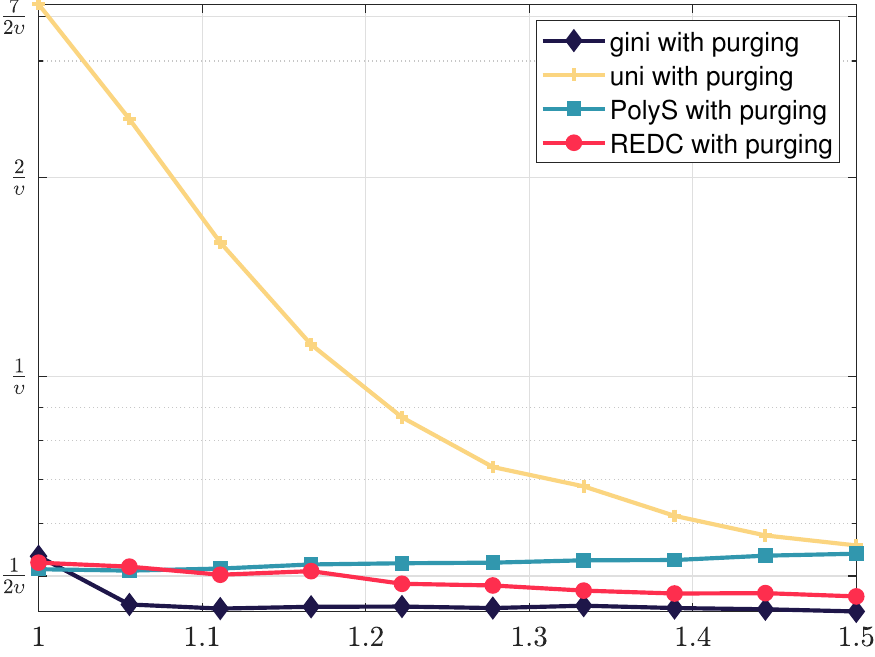}
	\caption{Relationship between computational load and Job execution time.Another presentation of the data in Figure.\ref{result5:sub2}}
	\label{result6}
\end{figure}

To further analyze the delay variation with \(\Gamma\), Figure.\ref{result6} illustrates the trends for four solutions. With additional signaling overhead, we manage to compress latency, bringing our scheme closer to the ideal scenario. However, our low-redundancy performance falls short of PolyS, attributable to the low-code-length redundancy \(\epsilon_{\mathrm{dec}}\) of the fountain code.

The future deployment of these methods within real edge networks holds significant promise. By intelligently combining computing tasks with transmission characteristics, it is possible to fully exploit the system's potential for multiplex transmission and parallel computing. Achieving this, however, will necessitate the development and implementation of more flexible and nuanced strategies.
\section{Conclusions and Future Directions}
In the pursuit of low-latency, high-reliability computing offloading services, our study considers the random arrival of workload jobs. Accordingly, we apply rateless coding to tasks, adaptively generating encoded CUs. At the onset of offloading, valid nodes within the edge network are identified and used as scheduler nodes for queuing theory modeling. This model is continuously updated based on the returned states from the nodes, from which a weight vector for task offloading is derived. Subsequently, computational tasks are dispatched in accordance with computing power. Our framework, REDC, navigates the complexities of edge network nodes. It takes into account their heterogeneity, unstable computational statuses, the volatility of the access environment, and the unpredictability of failures. It does so by maximizing the utilization of node computational resources, thereby minimizing the delay of sequential task execution.

There is room for further optimization in our strategy, particularly concerning the selection of degree distribution. The creation of the overall strategy can leverage a reinforcement learning scheme, enhancing flexibility. As edge computing offloading can potentially involve user-sensitive data, future research should take into account privacy-preserving encoding and offloading computation. Moreover, the potential presence of malicious nodes within the edge network necessitates the consideration of the security of computing results.

%\begin{thebibliography}{1}
% references section

\begin{appendices}
	\section{Density Evolution}
\subsubsection{Input Symbols Degree Distribution}
The encoding procedure of fountain codes include the independent selection of input symbols, utilized in generating output symbols (or encoding symbols) adhering to degree $\Omega (x) = \sum_d {\Omega _d}{x^d}$. Given that the count of input symbols is represented by K, the probability of a encoding edge being connected to an input symbol is 1/K. Thus, the probability distribution of the degrees of input nodes is typically characterized by a binomial distribution:
\begin{equation}\label{Lambda}
	{\Lambda _d} = \left( {\begin{array}{c}
			{\beta \mathcal{N}}\\
			d
	\end{array}} \right){P^d} - P^{\beta \mathcal{N} - d}
\end{equation}
$\beta$ represents the mean of ${\Omega _d}$, where  $\beta = {\Omega ^\prime }(1)$, and $P = 1/K$. As $ \mathcal{N} \to \infty $, the binomial distribution can be reasonably approximated by a Poisson distribution with parameter $\lambda = \alpha $. Thus, ${\Lambda _d} = \frac{{(\beta\mathcal{N} /K)\exp ( - \beta\mathcal{N} /K)}}{{d!}} = \frac{{e^{- \alpha} {\alpha ^d}}}{{d!}}$.

Introducing $\alpha $ as the average degree of the input symbols, we can derive that $\alpha = \beta (1 + \epsilon )$, where $\epsilon = (\mathcal N- K)/K$. This enables us to express the degree function of the input symbols as:
\begin{equation}
	\begin{split}
		\Lambda (x) &= \sum\limits_d {} {\Lambda _d}{x^d}  = \sum\limits_d {} \frac{{{e^{ - \alpha }}{\alpha ^d}}}{{d!}}{x^d} \\
		& ={e^{ - \alpha }}{e^{\alpha x}} = \exp ( - \alpha (1 - x))
	\end{split}
\end{equation}

\subsubsection{Density Evolution and Computation Graph}
Density evolution is a powerful tool for analyzing the performance of fountain codes in terms of their probability changes during the belief propagation (BP) decoding process. This methodology thereby assists in the evaluation of the asymptotic performance of these codes and the determination of their convergence properties. A commonly employed method to analyze fountain codes in erasure channels is the "and-or tree" analysis \cite{1998Analysis}.
In the Tanner graph of fountain codes, the "or" and "and" designations are respectively given to the input and output nodes. The BP decoding process is conceptualized as a sequence of alternating iterations, with "and" and "or" nodes continuously exchanging the roles of parent nodes.

In conceptualizing distributed coded computing as a data transmission problem over a binary erasure channel, we regard the probability of encountering a straggler node as $\epsilon$. We employ density evolution to analyze the corresponding probabilities, and subsequently optimize the encoding scheme to mitigate the impact of such straggler nodes.

In the and-or tree analysis, a node value of 0 signifies the node's unrecovered state. Let's denote the probability of a leaf node being 0 as $\delta$, assuming that the probabilities associated with different nodes are independent. The probability of the root node of the tree, denoted as $\mathcal{T}_l$, being 0 can thus be expressed as $y_l$. The and-or tree exhibits a threshold property, given by ${\delta}^{\mathrm{th}}$. If $\delta>{\delta}^{\mathrm{th}}$, $y_l \to 1$ as $l \to \infty $. Conversely, for $\delta<{\delta}^{\mathrm{th}}$, $y_l \to 0$ as $l \to \infty $.
The selection probability $\{\lambda_i\}$ corresponds to the input degree distribution from an edge perspective, representing the probability that an "or" node selects $i$ child nodes for an OR operation. Similarly, the selection probability $\{\omega_i\}$ corresponds to the output degree distribution from an edge perspective, representing the probability that an "and" node selects $i$ child nodes for an AND operation. Additionally, we define two non-negative values, $a$ and $b$, which respectively represent the probability of the initial value of an "or" node being 0 and the probability of the initial value of an "and" node being 1.

The following lemma may be established, for all $l \ge 1$:
\begin{equation}
	\begin{split}
		\lambda (x) = \Lambda '(x)/\Lambda '(1) = \sum\nolimits_{d = 1}^K {{\lambda _d}} {x^{d - 1}}\\ 
		\omega (x) = \Omega '(x)/\Omega '(1) = \sum\nolimits_{d = 1}^K {{\omega _d}} {x^{d - 1}}
	\end{split}
\end{equation}

The terms $\Lambda(x)$ and $\Omega(x)$ correspond to the degree distributions of the input and output symbols respectively. By applying these expressions, one can derive the degree distributions from the edge perspective for both the input and output symbols.

\begin{equation}\label{DensityFx}
	\begin{split}
		f(x) &= a\lambda (1 - b\omega (1 - x))\\
		{y_l} &= f({y_{l - 1}})
	\end{split}
\end{equation}

Analytical Perspective: Consider an "or" node at an even depth. The probability that the root node of $\mathcal{T}_{l-1}$ is 0 is denoted as ${y_{l - 1}}$. 
Now consider an "and" node at an odd depth. This node will yield a value of 1 only if all of its child nodes are also 1. Therefore, we can express ${x_l}$ as the sum of ${x_{l,i}}$, giving us ${x_l} = \sum\nolimits_i{x_{l,i}} = \sum\nolimits_i \omega {(1 - {y_{l - 1}})^i} = \omega (1 - {y_{l - 1}})$.
Similarly, for an "or" node, it can hold a value of 0 only when all of its child nodes are 0. These child nodes follow the distribution $\lambda (x)$, implying ${y_{l,i}} = a{\lambda_i}{(1 - {x_l})^i}$, and ${y_l} = a\lambda (1 - b\omega (1 - {y{l - 1}}))$.

Consequently, the degree distributions from the edge perspective are denoted as $\omega (x)$ and $\lambda (x)$ respectively. If we set ${y_0} = \delta$, the probability that the input symbol is not restored can be defined by the following system of equations:

\begin{equation}\label{DensityYl}
	\left\{ {\begin{array}{l}
			{{y_0} = \delta }\\
			{{y_l} = \lambda (1 - \omega (1 - {y_{l - 1}}))}
	\end{array}} \right.
\end{equation}

As expressed in Eq.\ref{DensityFx}, let's set $a=\varepsilon$ and $b=1$. Subsequently, we define $\varepsilon ^{\mathrm{th}}$ as the threshold of the AND-OR tree, implying that $1-\varepsilon ^{\mathrm{th}}$ serves as the recovery threshold probability for the LT code. The dichotomy method can be employed to determine the value of $\epsilon ^{\mathrm{th}}$. The procedure is briefly outlined as follows: initially, let $\epsilon$ be a constant $c$ within the interval $[0, 1]$. If $y_l \to 0$ as $ l \to \infty $, then let $ \varepsilon = \frac{\varepsilon}{2}$. Otherwise, assign $ \varepsilon = \frac{\varepsilon+1}{2}$. This iterative process continues until the solution's precision meets the set error threshold $\varepsilon$.

The accurate determination of $\varepsilon ^{\mathrm{th}}$ via the density evolution equation\ref{DensityYl} hinges critically on two conditions \cite{2020Factored}: 1) node degree values in the computational graph are independent, and 2) the graph forms a tree.

\end{appendices}

\bibliography{reference}
\bibliographystyle{IEEEtran}
%\end{thebibliography}

\end{document}